\documentclass[a4paper,10pt]{article}
\usepackage{amsmath}
\usepackage{amssymb}
\usepackage{amsthm}
\usepackage{mathrsfs}
\usepackage{bm}
\usepackage[UKenglish]{babel}
\usepackage{csquotes}
\usepackage[top=2cm,bottom=3cm,left=3cm,right=2cm]{geometry}
\usepackage{enumerate}
\usepackage{slashed}
\usepackage{marvosym}
\usepackage{graphicx}
\usepackage{setspace}
\onehalfspace
\usepackage[backend=bibtex, style=numeric, sorting=none]{biblatex}
\renewbibmacro{in:}{%
  \ifentrytype{article}{}{\printtext{\bibstring{in}\intitlepunct}}}
\bibliography{NeutronStars}
\newcommand{\upd}{\mathrm{d}}

\allowdisplaybreaks[1]
\usepackage{hyperref}

\allowdisplaybreaks[1]

\newtheorem{theorem}{Theorem}[section]
\newtheorem{lemma}[theorem]{Lemma}

\newtheorem{corollary}[theorem]{Corollary}

\newtheorem{sublemma}{Sublemma}[theorem]

\begin{document}

\title{Slowly Decaying Waves on Spherically Symmetric Spacetimes and Ultracompact Neutron Stars}
\author{Joe Keir \\ \\
{\small DAMTP, Centre for Mathematical Sciences, University of Cambridge}, \\
{\small \sl Wilberforce Road, Cambridge CB3 0WA, UK} \\ \\
\small{j.keir@damtp.cam.ac.uk}}
\maketitle

\begin{abstract}
 We prove that, in a class of spherically symmetric spacetimes exhibiting stable trapping of null geodesics, linear waves cannot (uniformly) decay faster than logarithmically. When these linear waves are treated as a model for nonlinear perturbations, this slow decay is highly suggestive of nonlinear instability. We also prove that, in a large class of asymptotically flat, spherically symmetric spacetimes, logarithmic decay actually holds as a uniform upper bound. In the presence of stable trapping, this result is therefore the best one can obtain. In addition, we provide an application of these results to \emph{ultracompact neutron stars}, suggesting that all stars with $r < 3M$ might be unstable.
\end{abstract}

\tableofcontents

\section{Introduction}
In recent years a great deal of progress has been made in the study of issues of stability in General Relativity. The Einstein field equations can be written (once a suitable gauge has been fixed) as
\begin{equation}
\Box_g g_{\mu\nu} = \mathcal{N}_{\mu\nu}(g, \partial g) + T_{\mu\nu}
\end{equation}
where $\Box_g = g^{\mu\nu}\partial_\mu \partial_\nu$ is the top order terms in the (scalar) Laplace-Beltrami operator associated to the metric $g$, $\mathcal{N}$ contains nonlinear terms, and $T_{\mu\nu}$ depends on the matter fields present, as well as the metric. As a starting point for the study of this problem, it is therefore appropriate to study \emph{linear wave equations} on a given background spacetime. The hope is that, once this problem is sufficiently well understood, insights can eventually be carried over to the full nonlinear problem. See, for example, \cite{Dafermos2013} for discussion of linear wave equations on various backgrounds. Note, however, that outside of the monumental work of Christodoulou-Klainerman \cite{Christodoulou1993}, few fully nonlinear results are available. With the idea of eventually addressing the asymptotic stability of the nonlinear problem in mind, we would like to understand whether these waves decay, and if so, whether they generically decay sufficiently quickly for nonlinear applications.

The main geometric obstacles to the decay of waves are \emph{superradiance} and \emph{trapping}, and in this paper we will only be concerned with the latter. ``Trapping'' is the phenomenon whereby null geodesics remain in some bounded region forever. Since waves can be localised near null geodesics this clearly provides an obstruction to decay. Indeed, \cite{Sbierski2013a} shows that a certain kind of estimate linked to decay, known as an ``integrated local energy decay '' estimate \emph{cannot} hold whenever trapping is present (regardless of whether the trapping is stable or unstable). On the other hand, in most applications, this trapping is of an unstable (sometimes referred to as ``normally hyperbolic'') character, and since (by the uncertainty principle) waves cannot be \emph{entirely} localised on a null geodesic, one can overcome this difficulty, usually by appealing to a ``higher order energy''. ``Superradiance'' is the phenomenon whereby a suitably defined energy can fail to be positive-definite in some region, leading to the amplification of certain frequency ranges. This will not be an issue for us, as we will be concerned with spacetimes which do not exhibit superradiance.

In this paper we will be concerned with the situation in which trapping occurs, but is \emph{not} normally hyperbolic. Instead, the spacetimes we are interested in exhibit \emph{stable trapping}, which means that an open set of null geodesics are trapped. As one might expect, this leads to much slower decay rates. In a recent pair of papers (\cite{Holzegel2013}, \cite{Holzegel2014}), Holzegel and Smulevici examined a similar situation\footnote{Note that, in Kerr-AdS, the spacetime studied in \cite{Holzegel2013} and \cite{Holzegel2014}, the trapped null geodesics are ``trapped'' between some finite radius and the AdS boundary, which is not, technically, a bounded region, although for many purposes it can be treated as such.}, and found that linear waves only decay \emph{logarithmically}, that is, at a rate $(\log\tau)^{-2}$, where $\tau$ labels the leaves of an appropriate foliation. This can be contrasted with the asymptotically flat case, in which one typically obtains $\tau^{-2}$ decay. In this paper, for spherically symmetric spacetimes, we show that these results are in fact generic: whenever a spherically symmetric spacetime exhibits stable trapping, logarithmic decay of linear waves is the best one can hope for. Note that \cite{Moschidis:2015wya} obtained logarithmic decay on a very general class of backgrounds, demonstrating that in many cases (even when one expects \emph{nonlinear} instability), \emph{linear} waves will still decay at least logarithmically.

Holzegel and Smulevici were able to prove logarithmic decay as both an upper bound \cite{Holzegel2013} and a lower bound \cite{Holzegel2014} for the uniform decay of linear waves on Kerr-AdS, and we shall reproduce both of these results in our context. From the point of view of the nonlinear problem, it is the lower bound which is most important - as already mentioned, the idea behind studying the linear problem is to prove sufficiently fast decay to be able to obtain decay for the nonlinear problem. Whilst $\tau^{-2}$ decay \emph{is} sufficiently fast for this purpose, logarithmic decay is not, and so it is suggestive of nonlinear instability\footnote{See \cite{Dias2012a}, however, for an argument in favour of \emph{stability}, even in the case of very slow decay.}. It is, however, far from a \emph{proof} of nonlinear instability - we expect that, if \emph{generic} nonlinearities are included, then the system would exhibit an instability. However, the nonlinearities in the Einstein equations are far from generic, and indeed exhibit many remarkable features (such as a version of the \emph{null condition} of \cite{Klainerman1986}), so it is not impossible that the nonlinear system is stable, although it does appear unlikely.

Regarding the \emph{nature} of the instability, the upper bound is very informative. This tells us that linear waves really \emph{do} decay, albeit at a slow rate. Thus, if an instability is present, then it must be a truly nonlinear effect, which could not be detected by looking for ``growing mode'' solutions to the linearised equations, for example. This suggests that, initially, perturbations to the solution really would decay, but nonlinear effects would cause sufficient ``clumping'' of the perturbation to cause gravitational collapse before the perturbation completely decays. This can be compared with the conjectured instability of AdS space (\cite{DafermosUnpublished}, \cite{Bizon2011}) In view of the upper bound which we will prove, this is the nature of the instability which we propose.

A point which deserves some emphasis is the \emph{generality} of our results. Our results apply to the linear wave equation $\Box_g\phi + F\phi = 0$, where $F$ is an \emph{arbitrary} function of the radial co-ordinate$r$. Our lower bound requires nothing more than spherical symmetry and stable trapping, and does not depend on either the asymptotic or topological properties of the spacetime. Our upper bound requires a little more: in addition to spherical symmetry we require asymptotic flatness and a certain fall off rate for the function $F$, as well as trivial topology for the spacetime. However, it is likely that both spherical symmetry and this topological condition can be relaxed in the case of the upper bound. It is interesting that logarithmic decay can be proven under such general conditions (c.f.\ \cite{Burq1998}), although, as previously mentioned, this is not very useful for nonlinear applications.

One interesting application of these ideas is to the proposed astrophysical objects known as \emph{ultracompact neutron stars}. Essentially, these are fluid objects with radii less than around $3M$, defined by the property of possessing a \emph{photon sphere}. Of course, no such object can exist with radius less than $2M$ and, by Buchdahl's theorem \cite{Buchdahl1959}, neither can any such object with radius less than $(9/4)M$. Our results suggest that, in fact, all such objects are unstable to gravitational collapse. In particular, this includes \emph{all} spherically symmetric fluid objects with radius less than $3M$, and suggests the new theoretical bound $r > 3M$ on the radius of such stars, independent of their particular properties.

We first turn to a more in-depth discussion of these ultracompact stars, including a demonstration that our results are actually applicable to the linearised equations of motion for such objects. After this follows the more mathematical parts of the paper, in which we first prove the lower bound on decay of linear waves, followed by the upper bound.

\section{Ultracompact Neutron Stars}
\label{neutron star section}

\subsection{Background}
In this section we discuss the application of our results to the case of \emph{ultracompact neutron stars}. As mentioned above, these are astrophysical objects whose radius is $\lesssim 3M$, where $M$ is the mass of the star. The significance of this radius is that, for non-rotating stars, a \emph{photon sphere} is present at $r = 3M$. This is a surface upon which circular null orbits are possible, i.e. it is a surface exhibiting trapping, although this trapping will typically be \emph{unstable}. As such, trapping occurs at a local \emph{maximum} of the effective potential. Generally, we would include stars with a radius slightly larger than $3M$ in our definition of ``ultracompact'', if they still possess a photon sphere (which would now be inside the star - see \cite{Karlovini2002}).

The crucial difference between these stars and a Schwarzschild black hole (which also possesses a photon sphere) is that these geometries do not possess a horizon, and are completely regular down to $r = 0$. As such, when a null geodesic possesses nonzero angular momentum, it will feel the usual angular momentum barrier near $r=0$, meaning that the effective potential will diverge as $r^{-2}$. However, as we have just observed, there is a local maximum of the effective potential located at $r = 3M$, implying the existence of a \emph{local minimum} at some point in between. We can compare this with the effective potential in the Schwarzschild geometry, which does not have an effective minimum, and instead decreases monotonically behind the photon sphere. Thus Schwarzschild black holes do not exhibit \emph{stable} trapping, though ultracompact stars do.

The stable trapping present in ultracompact stars has been known of for some time - see \cite{Detweiler1975}. Comments on the effect of this stable trapping on the trajectories of massless particles were made in \cite{Kembhavi1980}, and it was noted \cite{Chandrasekhar1991} that this can lead to slowly decaying waves, including certain metric perturbations. However, to the author's knowledge, our work provides the first \emph{quantitative} estimates on the rate of decay of waves on ultracompact star backgrounds, as well as the first indication that \emph{non-rotating} ultracompact stars might be unstable to gravitational collapse. See \cite{Andersson1996}, \cite{Kokkotas1999} and \cite{Nemiroff1993} for further details of the properties of these objects.

It should be noted that a linear instability of spacetimes with an ergoregion but no horizon was shown in \cite{Friedman1978}, and this was applied to rapidly rotating ultracompact stars in \cite{Comins211}, and numerically investigated by various authors (see \cite{1996MNRAS.282..580Y}, \cite{Kokkotas2004} and \cite{Brito:2015oca} for an overview). This instability is of a qualitatively different character to the one we will propose, though they are in some sense connected. The idea behind the instability proposed in \cite{Friedman1978} is that sufficiently rapidly rotating ultracompact stars will develop an ergoregion, which may overlap with the trapped region. Waves may then become simultaneously trapped and amplified by the superradiant phenomenon, leading to an instability. In fact, it is considered to be a key observation that this \emph{does not} occur in the Kerr spacetime, where the trapped regime and the superradiant regime are in a certain sense disjoint \cite{Dafermos2011}. On the other hand, for rotating ultracompact stars, stable trapping leads to a much larger class of null geodesics being trapped - indeed, an \emph{open} set in the space of null geodesics is now trapped, and so one might expect an instability due to the overlapping of the trapped and superradiant regimes. This was indeed found (numerically) in \cite{Kokkotas2004}.

\subsection{Applicability of Our Results}
We would like to apply our results to the case of a non-rotating ultracompact star. For a linear test field, this follows immediately - the effective potential has a local minimum, as already discussed, and so we obtain logarithmic decay as a lower bound on the uniform decay rate. However, it is not clear that this really tells us anything about the stability of these objects. In vacuum, the behaviour of the wave equation is strongly linked to that of the metric, since the Einstein equations can be viewed as a system of quasilinear wave equations. In contrast, in the presence of matter, these are coupled to the field equations for the matter, which may lead to different behaviour.

Even at the linear level it is not clear that the behaviour of a linear scalar field has anything to do with metric perturbations when matter is present. Indeed, the linearised equations of motion for the metric are a set of linear wave equations for the metric perturbations, but \emph{coupled to the perturbations of the matter variables}, which we would prefer to avoid considering. This is particularly troubling as we would expect the local minimum of the effective potential to sometimes appear \emph{inside} the star, where we certainly cannot neglect the presence of matter.

Luckily, some of the metric perturbations decouple completely from the matter perturbations. These are the \emph{w-modes} of \cite{Kokkotas1992}, or the \emph{axial perturbations} of \cite{Chandrasekhar1991}. They are perturbations which transform as vectors under the $SO(3)$ spherical symmetry group. The equation satisfied by these perturbations can be found in \cite{Chandrasekhar1991}: it is of the form of a linear wave equation, which, after separating variables, gives an effective potential which is regular everywhere away from the origin, and possesses a local minimum. In fact, for a perfect fluid, the effective potential\footnote{Note that this does \emph{not} require a barotropic equation of state} is given by:
\begin{equation}
V(r) = \frac{e^{2v(r)}}{r^3}\left(r + h^2\left(r^3(\varepsilon(r) - p(r)) - 6m(r)\right)\right)
\end{equation}
where $\varepsilon(r)$ and $p(r)$ are the energy density and pressure of the fluid, $m(r) = \int_0^r \hat{r}^2 \varepsilon(\hat{r})\upd \hat{r}$ is the mass in the ball of radius $r$, and the function $v(r)$ is determined by
\begin{equation}
\frac{\upd v}{\upd r} = - \frac{1}{\varepsilon(r) + p(r)}\frac{\upd p}{\upd r}
\end{equation}
with the boundary condition that $e^{2v(R)} = 1 - 2M/R$ at the edge of the star $(r=R)$, where $M = m(R)$. Together with an equation of state for the fluid, this determines the effective potential.

Thus our results \emph{are} applicable, and indeed can be applied straightforwardly to these ``w-modes''. We conclude:
\begin{corollary}
No uniform decay estimate with faster than logarithmic decay can hold for axial perturbations of ultracompact stars. 
\end{corollary}

\subsection{Implications for the Stability of Ultracompact Stars}
Our results are, of course, compatible with stability of ultracompact stars, since we have not shown that perturbations grow, or even that they do not decay, but only that a uniform decay rate faster than logarithmic decay cannot hold. In fact, in section \ref{upper bound section}, we will show that linear waves do indeed decay at least logarithmically fast. However, when proving \emph{nonlinear} stability, one usually requires uniform decay on the linear level which is faster than logarithmic (and indeed, faster than $1/\tau$). Thus the usual methods could not be employed without significant changes, if one wished to prove the nonlinear stability of ultracompact stars.

There is reason to believe that the situation is actually worse than this. If a perturbation \emph{does not} decay on the linear level, then we would expect that, at the nonlinear level, generic perturbations would eventually get sufficiently ``concentrated'' in some region to form a trapped surface, after which a black hole would form\footnote{c.f.\ the proposed instability of AdS \cite{Bizon2011}}. If, instead, the perturbation decays but at an \emph{extremely} slow rate, we would expect the same picture to hold. In the light of results such as \cite{Burq1998}, logarithmic decay is the slowest decay one can reasonably expect whenever a uniform boundedness statement holds\footnote{It is trivial to show that such a statement holds for linear waves on ultracompact star backgrounds, due to the presence of a globally timelike Killing vector field.}. Thus, we \emph{tentatively} propose that ultracompact stars may be unstable.

We emphasise here the difference between our approach and some other studies of (linear) stability, which examine decay rates for individual (quasinormal) modes. It may well be the case that all such quasinormal modes decay - indeed, we would expect it. It is even the case that a uniform bound on the decay of linear perturbations can be proven (see section \ref{upper bound section}) - \emph{but not with any rate faster than logarithmic}, even though individual modes may all decay faster than this. For the non-linear problem, it is \emph{uniform} bounds which are important, and so examining individual modes may not capture the essence of the problem, as in this case.

Note that our instability, if it exists, is expected to be of a very weak character. Indeed, generic nonlinear perturbations of an ultracompact neutron star would be expected to decay at first, but so slowly that a singularity would form before they have decayed sufficiently. Note also that the instability is associated with high angular frequencies, which feel trapping the most. Thus, for example, it could not be seen by studying spherically symmetric perturbations, or even the first few angular frequencies, or indeed any individual mode. Also, due to the non-linear nature of the instability, we cannot obtain information about the timescale of this instability, which is usually done by examining a single unstable linear mode, although we do expect it to be very slow. This means that, if there is some process in nature which can cause these stars to form, and if they form at a sufficiently fast rate, then they may still be observed despite being unstable.

It is worth discussing in a bit more detail the nature of this proposed instability. As already mentioned, it is, by its very nature, a \emph{nonlinear} effect, which could not be seen in linear studies (except to the extent that it is observed in this paper). Furthermore, perturbations with support up to arbitrarily high angular momentum have to be considered (note that generic perturbations satisfy this condition), although, importantly, the proportion of the energy of a perturbation in the high angular momentum range could be very small. Note also that, unlike the previously discovered instability \cite{Kokkotas2004} of rotating stars, the end point of our proposed instability is clearly gravitational collapse. The instability associated with the interaction of superradiance and trapping \emph{could} lead to a ``spinning down'' of the star, until the ergoregion shrinks to a sufficiently small size or disappears altogether. Alternatively, and more dramatically, it could lead to sufficient amplification of perturbations near to the star to form a trapped surface, resulting in gravitational collapse. On the other hand, the most likely endpoints of our proposed instability appear to be either gravitational collapse, or the shedding of mass from the star, if our picture of the situation is accurate. Note that, if collapse occurs, it is expected to occur near to the photon sphere, and it may be the case that a small black hole forms there, but takes a very long time to absorb the rest of the neutron star.

Finally, we note that there is another reason to question the astrophysical ramifications of these ideas, even if the proposed instability is present in ultracompact stars. The reason for this is that the equation of state of realistic compact objects may already rule out the existence of ultracompact stars \cite{Iyer1985}, at least when we restrict to stars with $R \leq 3M$. Note, however, that stars with radii $R>3M$ but still possessing a ``photon sphere'' may be far more realistic from the point of view of the equation of state - see \cite{Karlovini2002}. Indeed, \cite{Andrade2001} found that it was not very difficult to construct neutron stars with first order phase transitions which possessed trapped w-modes (though the equations of state they considered were also not entirely realistic). In addition, there are some speculative proposals for equations of state admitting ultracompact solutions (e.g.\ \cite{Bahcall1989}), in the sense of $R \leq 3M$, and anisotropic neutron stars may become ultracompact much more easily \cite{Silva:2014fca}. Likewise, boson stars may be ultracompact (see, for example, \cite{Macedo:2013jja}), and, even more speculatively, gravastars might have any degree of compactness.

If any of these proposals are taken seriously, then our instability may be relevant. Indeed, all of these objects have been proposed at various times as possible alternatives to black holes, in the sense that certain astrophysical observations may not be able to tell the difference between these objects and black holes. It had already been established that, even if such objects exists, rapidly rotating ultracompact stars would be expected to be unstable. Our results suggest that even non-rotating ultracompact stars might be unstable to gravitational collapse, and furthermore this statement is \emph{independent} of the peculiar features of these stars - in particular, it is independent of any proposed equation of state which might give rise to such stars.

Other than their extreme compactness, there are other observations which may provide evidence in favour of a given astrophysical object being a black hole. It is often claimed that ringdown, such as that observed in the recent work of the LIGO group \cite{Abbott:2016blz}, provides indirect evidence for the presence of a horizon, although it has also been shown \cite{Cardoso:2016rao} that other (horizonless) ultracompact objects may give rise to similar gravitational wave signals. Thus, the results of this paper may play an important role in providing \emph{theoretical} evidence in favour of horizons.


Finally, we note that the paper \cite{Cardoso:2014sna}, which appeared after a preprint of this work, studied ultracompact stars numerically and found close agreement with the results presented here. This suggests that the slow decay of linear waves on these backgrounds, proved in this paper, may indeed give rise to nonlinear instabilities, as we have speculated.

This concludes the ``physics'' part of the paper. The remaining sections are more mathematical in nature, and concern first the construction of quasimodes, and the resulting proof of logarithmic decay as a lower bound. Afterwards, we prove the (sharp) result that logarithmic decay in fact does hold as a uniform \emph{upper bound}, under some slightly stronger conditions.

\section{Preliminaries}

\subsection{Conventions}
We will write $r$ for the area radius function, and $r^*$ will denote a suitably defined ``tortoise'' co-ordinate. Primes will denote differentiation with respect to $r^*$, so, for example, $u' = \partial_{r^*}u$.

The function satisfying an appropriate wave equation will be called $\phi$, and we define the rescaled quantity $\psi = r\phi$.

When an integral is defined over a geometric surface, the induced volume form will be omitted. For example, if $\Sigma_t$ is a surface of constant $t$ in Minkowski space, then
\begin{equation}
\int_{\Sigma_t} u^2 = \int_{r = 0}^\infty \upd r\int_0^\pi \upd\theta \int_0^{2\pi}\upd\varphi \, r^2 \sin\theta u^2
\end{equation}
Such integrals will usually involve the quantity $\phi$, and it is particularly important to remember the weight $r^2$ which will appear in all such integrals.

By contrast, integrals involving the quantity $\psi$ will usually have the volume form explicitly displayed, and so these do not have any hidden $r$ weights. We will also use, for the standard volume form on the unit sphere
\begin{equation}
\upd\Omega = \sin\theta \upd\theta \, \upd\varphi
\end{equation}

Angular derivatives will be written with a slash, so, for example, $\slashed{\nabla}\phi$ denotes the set of angular derivatives of $\phi$. It is important to remember that these are derivatives on the sphere of radius $r$, so that, written in terms of he co-ordinate derivatives $\partial_\theta$ and $\partial_\varphi$, the ``slashed'' quantities involve a weight of $r^{-1}$. We will also write $\Omega_i$ for an element of the standard basis for the Lie algebra of $SO(3)$, which are sometimes called ``angular momentum operators''. Note that these operators should be considered as containing a weight $r$ (for example, we can take $\Omega_3 = \partial_\varphi$, then $||\Omega_3|| = r\sin\theta$).

Finally, our conventions allow us to redefine the constant $C$ without explicitly renaming it. Also, the notation $ a \lesssim b$ means that there exists some positive constant $C$ (independent of all the relevant parameters, otherwise we would write it out explicitly) such that $a \leq Cb$. Similarly, the notation $a \sim b$ means that there exist positive constants $c, C$ such that $ca \leq b \leq Ca$.

\subsection{The Metric and Co-ordinates}
We consider the spherically symmetric metric
\begin{equation}
\upd s^2 = - a(r) \upd t^2 + b(r) \upd r^2 + r^2 \upd \sigma_{\mathbb{S}^2}
\label{metric}
\end{equation}
where $\upd \sigma_{\mathbb{S}^2}$ is the standard metric on the sphere, and $a$ and $b$ are smooth. We will also frequently use the tortoise-like co-ordinate $r^*$, defined by
\begin{equation}
\frac{\upd r^*}{\upd r} = b^{\frac{1}{2}}a^{-\frac{1}{2}}
\label{tortoise}
\end{equation}
in terms of which the metric takes the form
\begin{equation}
\upd s^2 = -a \upd t^2 + a\upd r^{*2} + r^2 \left( \upd \theta^2 + \sin^2 \theta \, \upd\varphi^2 \right)
\end{equation}
We are interested in the case where the function (which will later turn out to be the effective potential for null geodesics)
\begin{equation}
V_{\textnormal{null}}(r) := \frac{1}{r^2} a(r)
\end{equation}
has a local minimum $V_{\textnormal{min}}$ at some $r = r_{\textnormal{min}}$. For proving our lower bound, we only require the metric to be regular in these co-ordinates in some open region $U$, such that $r_{\textnormal{min}} \in U$. On the other hand, proving logarithmic decay as an upper bound is an inherently global affair, and so we will require the above co-ordinates to be global (up to the usual issues on the sphere).

We also require the metric to be asymptotically flat, although we should note that this will only be required in order to prove our upper bound - the lower bound only depends on the local geometry around the trapped geodesics. The precise (weak) statement of asymptotic flatness which we require is the following:
\begin{equation}
\begin{split}
a &= 1 + a_1 r^{-1} + \mathcal{O}(r^{-1-\epsilon_a}) \text{\quad, } \epsilon_a >0 \\
b &= 1 + b_1 r^{-1} + \mathcal{O}(r^{-1-\epsilon_b}) \text{\quad, } \epsilon_b >0
\label{asymptotic flat}
\end{split}
\end{equation}
for some constants $a_1$ and $b_1$. In order to recover pointwise bounds, we also require that the \emph{same} decay rates hold for the metric components $g_{\mu\nu}$ replaced by $Z^{(n)}g_{\mu\nu}$, where $Z\in\{\partial_t, \Omega_{ij}:= x^i\partial_j - x^j\partial_i\}$ is a vector field generating either time translation or spatial rotation, and $n$ is a multi-index, with $|n|\leq N$ for some sufficiently large $N$.

We will occasionally refer to the orthonormal basis:
\begin{align}
e^0 &= a^{\frac{1}{2}}\ \upd t & e^1 &= a^{\frac{1}{2}}\ \upd r^* \nonumber\\ 
e^2 &= r\ \upd \theta & e^3 &= r \sin\theta \ \upd\varphi
\end{align}
and its dual basis of vector fields $\{E^0, \ldots E^3 \}$. Finally, we will also label the co-ordinate vector field $T = \partial_t$, with respect to either the $(t,r,\theta,\varphi)$ or the $(t,r^*\theta\,\varphi)$ co-ordinates, as well as the vector field $Y = \partial_{r^*}$.

\subsection{Stable Trapping}
\label{stable trapping}
We are interested in the case where null geodesics are \emph{stably trapped}. One way to define this would be to require that all Jacobi fields along a certain null geodesic possess conjugate points, but since we are in spherical symmetry and geodesic flow is integrable we can do better. Without loss of generality, we can assume a null geodesic to lie in the equatorial plane; to have angular momentum $m$ and energy $\mathcal{E}$. Then the geodesic satisfies
\begin{equation}
\mathcal{E}^2 \frac{b}{a} \left(\frac{\upd r}{\upd t}\right)^2 + m^2 \frac{a}{r^2} = \mathcal{E}^2
\end{equation}

Defining a rescaled time co-ordinate $t = \mathcal{E} \tilde{t}$, the geodesic equation implies
\begin{equation}
\left(\frac{\upd r^*}{\upd t'}\right)^2 + m^2 \frac{a}{r^2} = \mathcal{E}^2
\end{equation}
and we see that our null geodesic experiences an effective potential
\begin{equation}
V(r) = m^2 \frac{a}{r^2} = m^2 V_{\textnormal{null}}(r)
\end{equation}
We say that \emph{trapping} occurs when $V_{\textnormal{null}}$ has a local extremum, and (local) \emph{stable trapping} when it has a local minimum. In this case, stable trapping means that all null geodesics near to the one which sits at the local minimum will remain in some bounded region containing the local minimum. As we will see, this phenomenon is closely related to very slow decay of certain solutions to the wave equation.

\subsection{The Wave Equation}
Next we will find the effective potential for the wave equation, after separating variables. The Laplace-Beltrami operator in the background (\ref{metric}) is
\begin{equation}
\Box_g \phi = -a^{-1} \partial_t^2 \phi + r^{-2} a^{-\frac{1}{2}}b^{-\frac{1}{2}} \partial_r\left(r^2 a^{\frac{1}{2}}b^{-\frac{1}{2}} \partial_r \phi\right) + \slashed{\triangle} \phi
\end{equation}
where $\slashed{\triangle}$ is the Laplacian on the sphere of radius $r$. Defining $\psi = r\phi$, and once again transforming to the radial `tortoise' co-ordinate (\ref{tortoise}), we find
\begin{equation}
ar \Box_g \phi = -\partial_t^2 \psi + \partial_{r^*}^2 \psi - r^{-1}a^{\frac{1}{2}}b^{-\frac{1}{2}}\partial_r \left(a^{\frac{1}{2}}b^{-\frac{1}{2}}\right) \psi + a \slashed{\triangle} \psi
\label{boxpsi}
\end{equation}

We are interested in capturing some of the features of perturbations to the metric (\ref{metric}). In the vacuum case, the wave equation captures many of the features of linear metric perturbations, however, we are also interested in the case where matter is present, for example in ultracompact neutron stars. Linear perturbations to these backgrounds will take the form of a wave-like equation for the metric perturbation, \emph{coupled} to some other equations for the matter perturbations. Nevertheless, often some subset of perturbations will decouple and satisfy an equation of the form:
\begin{equation}
\Box_g \phi + F(r) \phi = 0
\label{wave equation 1}
\end{equation}
as already mentioned in section \ref{neutron star section} this is in fact the case for ultracompact stars (see \cite{Chandrasekhar1991}), and in this case $\phi$ is just a metric perturbation, in a certain gauge. The function $F(r)$ is essentially playing the role of the (unperturbed) energy momentum tensor.\footnote{Note that we can also easily include terms linear in the first derivatives of $\phi$, but such terms can cause minor and unenlightening complications.} Thus, equation (\ref{wave equation 1}) is the one we will study, with the ``asymptotic freedom'' restriction on $F$ that, as $r \rightarrow \infty$, 
\begin{equation}
\begin{split}
F(r) &= \mathcal{O}(r^{-2 - \epsilon_F}) \text{\quad, } \epsilon_F > 0 \\
F'(r) &= \mathcal{O}(r^{-3 - \epsilon_F})
\end{split}
\label{asymptotic freedom}
\end{equation}
Finally, we also require the condition $F \geq 0$.

\subsection{The Reduced Equation}
\label{reduced section}
Using (\ref{wave equation 1}), expanding in spherical harmonics, and taking the Fourier transform in time so that
\begin{equation}
\psi(r, t, \theta, \phi) = \frac{1}{\sqrt{2\pi}}\int_\infty^\infty e^{-i\omega t}\sum_{m,\ell} P_{m \ell}(\cos \theta)e^{im\phi} u_{m\ell}(r)
\end{equation}
we find that the $u_{m\ell}$ satisfy
\begin{equation}
\partial_{r^*}^2 u_{m\ell} - \ell(\ell+1) V^{(\ell)}_{\textnormal{eff}}(r) = \omega^2 u_{m\ell}
\label{reduced1}
\end{equation}
where we have defined the \emph{effective potential}
\begin{equation}
V^{(\ell)}_{\textnormal{eff}}(r) = a r^{-2} + \frac{1}{\ell(\ell+1)}\left( a^{\frac{1}{2}} b^{-\frac{1}{2}} r^{-1} \partial_r(a^{\frac{1}{2}}b^{-\frac{1}{2}}) + aF \right)
\end{equation}
We now define a semiclassical parameter
\begin{equation}
h^{-2} = \ell(\ell+1) 
\end{equation}
so that we can alternatively write the effective potential as
\begin{equation}
V^{(h)}_{\textnormal{eff}}(r) = a r^{-2} + h^2 \left( a^{\frac{1}{2}} b^{-\frac{1}{2}} r^{-1} \partial_r(a^{\frac{1}{2}}b^{-\frac{1}{2}}) + aF \right)
\end{equation}
Finally, defining the eigenvalue $\kappa = \omega^2 h^2$, the reduced equation takes the Schr\"odinger-type form
\begin{equation}
-h^2 \partial_{r^*}^2 u_{m\ell} + V^{(h)}_{\textnormal{eff}}(r) u_{m\ell} = \kappa u_{m\ell}
\label{Schrodinger}
\end{equation}

Note that we have taken the Fourier transform in time, and, importantly, we are not really justified in assuming the Fourier transform exists for our solution. For the purpose of constructing quasimodes and proving our lower bound this is not an issue, since we will actually \emph{construct} approximate solutions, and the above can be interpreted merely as motivation for our definition of the effective potential. Later on, for the purpose of proving our upper bound, we will need to deal with the fact that the solution may not be assumed \emph{a priori} to lie in $L^2$ in time. As such, we will have to cut-off in time, and then deal with the errors incurred by this process in an appropriate way. Since it is not used in the proof of the lower bound, we postpone discussion of this cut-off to a later part of the paper.

\subsection{Hardy Inequalities}
In this section we collect various Hardy inequalities, which we will make use of in the rest of the paper.
\begin{lemma}
\emph{(Hardy Inequalities)}
\label{Hardy lemma}

Suppose $u(r_+) = 0$. Then, for all $n \geq 0$:
\begin{equation}
\int_0^{r_+} r^n a^{\frac{1}{2}}b^{-\frac{1}{2}}|u|^2 \upd r^* \leq \frac{4}{(1+n)^2}\int_0^{r_+} r^{n+2} a^{-\frac{1}{2}}b^{\frac{1}{2}}|u'|^2 \upd r^*
\end{equation}
where $u' =\upd u/ \upd r^*$.

Suppose also, $r^{-n+1}|u|^2 \rightarrow 0$ as $r \rightarrow \infty$. Then:
\begin{equation}
\int_{r_+}^\infty r^{-n} a^{\frac{1}{2}}b^{-\frac{1}{2}}|u|^2 \upd r^* \leq \frac{4}{(n-1)^2} \int_{r_+}^\infty r^{2-n} a^{-\frac{1}{2}} b^{\frac{1}{2}} |u'|^2 \upd r^*
\end{equation}
\end{lemma}
\emph{Proof.}
\begin{equation}
\begin{split}
\int_0^{r_+} r^n a^{\frac{1}{2}} b^{-\frac{1}{2}} |u|^2 \upd r^* &= \int_0^{r_+} \frac{\upd r}{\upd r^*} r^n |u|^2 \upd r^* \\
&= -\int_0^{r_+} \left( 2r^{n+1} \mathop{\mbox{Re}}(\bar{u}u') + n a^{\frac{1}{2}} b^{-\frac{1}{2}} r^n |u|^2 \right)\upd r^* \\
\Rightarrow (1+n)\int_0^{r_+} r^n a^{\frac{1}{2}} b^{-\frac{1}{2}} |u|^2 \upd r^* &\leq 2\left(\int_0^{r_+} r^n a^{\frac{1}{2}}b^{-\frac{1}{2}}|u|^2 \upd r^* \right)^{\frac{1}{2}} \cdot \left(\int_0^{r_+} r^{n+2} a^{-\frac{1}{2}} b^{\frac{1}{2}}|u'|^2 \right)^{\frac{1}{2}}
\end{split}
\end{equation}
where we have used Cauchy-Schwarz in the last line. The second inequality follows from a similar calculation.

\subsection{Nondegenerate Energy Boundedness}
Often, the first statement one wishes to prove when analysing the wave equation on a particular background is one of \emph{nondegenerate boundedness}. Typically, on a black hole background, this involves exploiting the celebrated \emph{red shift effect} \cite{Dafermos2009}. In our case, however, there is neither a horizon nor an ergosphere nor any other obstruction to the use of the na\"{\i}ve ``T-energy'' to construct a non-degenerate energy, and so this is what we shall do.

Define the \emph{energy momentum tensor} associated with the field $\phi$ as
\begin{equation}
Q_{\mu\nu}[\phi] = \partial_\mu \phi \partial_\nu \phi - \frac{1}{2}g_{\mu\nu}\left(\partial_\rho\phi \partial^\rho\phi - F \phi^2\right)
\end{equation}
Note that this satisfies
\begin{equation}
\nabla^\nu Q_{\mu\nu}[\phi] = \frac{1}{2}(\partial_\mu F) \phi^2
\end{equation}
Now, define the current associated with the vector field $T$
\begin{equation}
J^T[\phi](\cdot) = Q[\phi](T, \cdot)
\end{equation}
Since $F$ is independent of time, we find that this current is conserved:
\begin{equation}
\mathop{\mbox{div}} J^T[\phi] = 0
\end{equation}
and hence we can define the conserved $T$-energy (recall that $F \geq 0$):
\begin{equation}
\mathcal{E}^T[\phi](\tau) = \int_{\Sigma_\tau} J^T[\phi]n_{\Sigma_\tau} \sim \int_{\Sigma_\tau} \left((\partial_t \phi)^2 + (\partial_{r^*} \phi)^2 + |\slashed{\nabla}\phi|^2 + F \phi^2 \right)
\end{equation}
where $\Sigma_\tau$ is the surface $t = \tau$, which has unit normal $n_{\Sigma_\tau}$, and all integrals are with respect to the volume form associated with the metric $g$. This energy satisfies the conservation law
\begin{equation}
\mathcal{E}^T[\phi](\tau) = \mathcal{E}^T[\phi](0)
\label{boundedness}
\end{equation}

We also define the \emph{higher order energy}:
\begin{equation}
\mathcal{E}^T_2[\phi](\tau) = \mathcal{E}^T[\phi](\tau) + \mathcal{E}^T[T\phi] + \mathcal{E}^T[Y\phi] + \sum_i \mathcal{E}^T[\Omega_i\phi]
\end{equation}

\section{The Lower Bound}
\label{lower bound section}

In this section we will follow the structure of the argument presented in \cite{Holzegel2014} to prove logarithmic decay as a lower bound on the uniform decay of waves on ultracompact neutron stars. Actually, as the argument is very general, we only require the effective potential (defined in section \ref{reduced section}) to have a local minimum somewhere, and so the results of this section hold in a wide class of spherically symmetric spacetimes.

The approach used in this section can be summarised as follows. We will first construct solutions to the reduced equation, with artificial Dirichlet boundary conditions imposed near the local minimum of the effective potential, but in the ``classically forbidden region'' for solutions with the given energy. Then, we will smoothly cut off these solutions to produce an approximate solution on the whole spacetime - these will be the quasimodes. They are oscillatory, non-decaying, and are localised around the local minimum of the effective potential. Finally, we will quantify the error produced by the cutting off procedure, which will be sufficiently small to allow us to conclude logarithmic decay as a lower bound.

In order to make the presentation self-contained, all the steps of the proof will be presented below. However, this section follows \cite{Holzegel2014} closely, with only minor alterations in most places.

\subsection{Weyl's Law}

We are concerned with solving the eigenvalue problem (\ref{Schrodinger}) in a region around a local minimum of the effective potential, with Dirichlet boundary conditions imposed nearby. We would like to be able to freely specify some appropriate eigenvalue for the quasimodes, but since they are solutions to a Dirichlet problem in a bounded domain, we only obtain a countable number (for each $h$) of possible eigenvalues. We can, however, establish the existence of a sequence of solutions, with boundary conditions imposed near the minimum of $V_\textnormal{n}$, and with eigenvalues tending to our specified value as $h \rightarrow 0$.

As in \cite{Holzegel2014}, we will prove a stronger result than we actually need, namely a version of Weyl's law. This states that the number of eigenvalues $\kappa$ of the Dirichlet problem for (\ref{Schrodinger}) in some small neighbourhood of an appropriate value scales as $h^{-1}$, as $h \rightarrow 0$. In fact we only require a sequence of such eigenvalues to approach our chosen value as $h \rightarrow 0$.

First, we state a simple lemma which mostly serves to introduce some notation.
\begin{lemma}
Let the local minimum of $V_{\textnormal{null}}$ be $V_{\textnormal{min}}$, and let this be attained at the point $r = r_{\textnormal{min}}$. Pick some $c>0$ (independent of $h$) sufficiently small such that there exist some $r_-$ and $r_+$ satisfying
\begin{equation}
r_- < r_{\textnormal{min}} < r_+
\end{equation}
where $V_{\textnormal{null}}(r_-) = V_{\textnormal{null}}(r_+) = V_{\textnormal{min}} + c$, and such that there are no local maxima of $V_{\textnormal{null}}$ in $[r_-, r_+]$.

Next, pick some $E > V_{\textnormal{min}}$ such that $E-V_\textnormal{min} < c$. Then, for all sufficiently small constants $\delta, \delta' > 0$ we can find some constant $c'$ such that
\begin{equation}
|r_- - r| < \delta' \Rightarrow V_{\textnormal{null}}(r) - \kappa > c'
\end{equation}
for all $\kappa \in [E-\delta, E+\delta]$. We obtain the same result with $r_-$ replaced by $r_+$. Also, for all sufficiently small $h$ the same statement holds with $V_{\textnormal{null}}$ replaced by $V^{h}_{\textnormal{eff}}$, and moreover the constants $E$, $c$, $c'$, $\delta$ and $\delta'$ can be chosen independent of $h$, in the sense that the result will hold for all sufficiently small $h$.
\label{aux lemma 1}
\end{lemma}

\begin{figure}[ht]
\centering
\includegraphics[width = 0.8\linewidth, keepaspectratio]{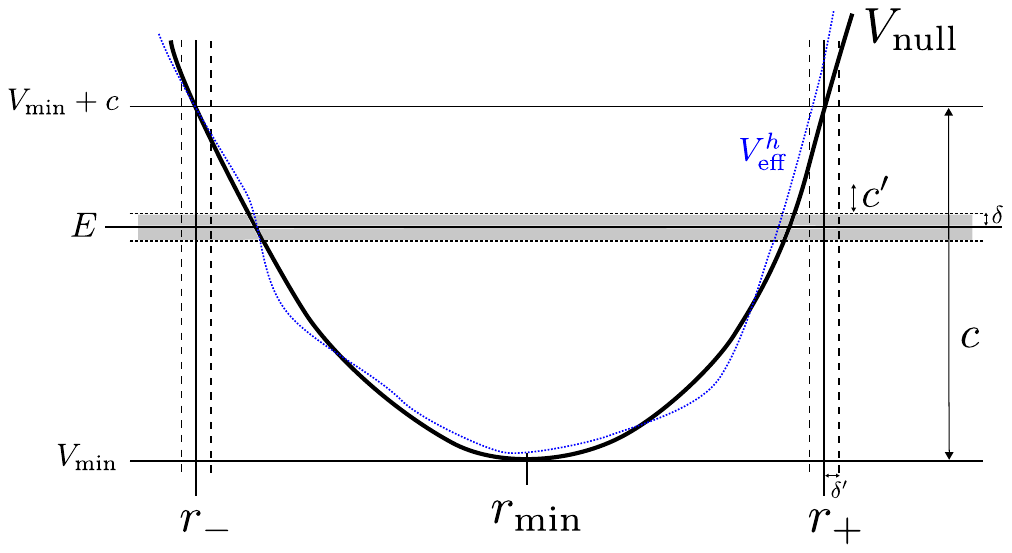}
\caption{A diagram illustrating lemma \ref{aux lemma 1}. The blue dotted curve is a graph of $V^h_{\textnormal{eff}}$ for some small value of $h$, so that its graph is close to that of $V_{\textnormal{null}}$. Note that the value of $c'$ illustrated is not the optimal value which could be chosen for $V_{\textnormal{null}}$, given the choices of $E$, $\delta$ and $\delta'$ illustrated, but some lower value. Such a choice ensures that the final statement of lemma \ref{aux lemma 1} is true, i.e.\ that the conclusions of the lemma remain true when $V_{\textnormal{null}}$ is replaced by $V^h_{\textnormal{eff}}$, for all sufficiently small $h$.}
\label{lemmafigure}
\end{figure}

\emph{Proof}. This is a simple consequence of the continuity of $V_{\textnormal{null}}$ and the uniform convergence of $V^{h}_{\textnormal{eff}}$ to $V_{\textnormal{null}}$. See figure \ref{lemmafigure}.
\\[5pt]
We will adopt the notation of \cite{Holzegel2014} and denote by $P_{DD}(r_1,r_2)$ the Dirichlet problem (\ref{Schrodinger}) with boundary conditions imposed at $r_1$ and $r_2$, by $P_{NN}(r_1,r_2)$ the corresponding Neumann problem, and by $P_{DN}(r_1,r_2)$ or $P_{ND}(r_1,r_2)$ the mixed Neumann/Dirichlet problems, with Dirichlet boundary conditions imposed at $r_1$ or at $r_2$ respectively.

We are now in a position to prove a version of Weyl's law.\footnote{Note that, in view of the compactness of the region $[r_-, r_+]$, we do not need to prove a version of the auxiliary lemma $4.1$ of \cite{Holzegel2014}, which is used to gain control over the number of eigenvalues in a non-compact region.}

\begin{lemma}
Consider the eigenvalue problem $P_{DD}(r_-, r_+)$. Fix an energy level $E > V_{\textnormal{min}}$ with $E - V_{\textnormal{min}}$ sufficiently small, and pick some $\delta < E-V_{\textnormal{min}}$ also sufficiently small, so that $(E + \delta) < c$, where $c$ was defined in lemma \ref{aux lemma 1}. Then, as $h\rightarrow 0$, the number of eigenvalues of $P_{DD}(r_-, r_+)$ in the interval $[E-\delta, E+\delta]$ satisfies
\begin{equation}
N[E-\delta, E+\delta] \sim \frac{1}{\pi h}\int_{r^*_-}^{r^*_+} \left(\sqrt{E - V(r^*)}\right) \chi_{\{V \leq E\}} \upd r^* 
\end{equation}
where $\chi$ is the indicator function.
\label{Weyl law}
\end{lemma}
\emph{Proof}. Equipartition the interval $[r_-, r_+]$ into $k$ intervals of length $\beta = (r_- - r_+)/k$, and consider the set of Dirichlet problems $P^i_D$ ($i=1,\ldots ,k$), where $P^i_D$ is the problem $P_{DD}(r_- + (i-1)\beta, r_+ + i \beta)$ but with $V^h_{\textnormal{eff}}(r)$ replaced by its \emph{maximum} value in the i\textsuperscript{th} interval, $V^i_+$. Also, let $P^i_N$ be the corresponding Neumann problem, but with $V^h_{\textnormal{eff}}(r)$ replaced by its \emph{minimum} value in the i\textsuperscript{th} interval, $V^i_-$. Finally, we donate by $N_{\leq E}(P)$ the number of eigenvalues of the problem $P$ which are less than or equal to $E$. Then we have the following lemma:
\begin{sublemma}
\label{sublemma on number of eigenvalues}
 The number of eigenvalues below the threshold $E$ satisfies
 \begin{equation}
 \sum_{i=1}^k N_{\leq E}(P^i_D) \leq N_{\leq E}(P_{DD}(r_-, r_+)) \leq \sum_{i=1}^k N_{\leq E}(P^i_N)
 \end{equation} 
\end{sublemma}

We shall prove lemma \ref{sublemma on number of eigenvalues} in two steps, first showing that the inequalities hold without modifying the potential $V_{\text{null}}$, and then showing that they continue to hold if the potential is modified as in the lemma. Note that we can characterise the smallest eigenvalue $\kappa_1$ to the problem $P_{DD}(r_-, r_+)$ by
\begin{equation}
 \kappa_1 = \inf_{f \in H^1_0([r_-, r_+]), \ ||f||_{L^2}\neq 0} \frac{\int_{r_-}^{r_+} ( h^2 |\partial_{r^*}f|^2 + V_{\text{eff}}^{(h)}(r) |f|^2 )\upd r^*}{||f||_{L^2}^2}
\end{equation}
where the space $H^1_0([r_-, r_+])$ can be defined as the completion in $H^1([r_-, r_+])$ of the $C^{2}$ functions on $[r_-, r_+]$ which vanish at the end-points. Similarly, by the minimax principle, we can characterise the $n$-th eigenvalue by
\begin{equation}
 \kappa_n = \inf_{\substack{(f_1, f_2, \ldots , f_n), \ f_i \in H^1_0([r_-, r_+]) \\ ||f_i||_{L^2}\neq 0, \ \langle f_i, f_j \rangle = 0 \ \forall i\neq j} } \max_{i \leq n} \frac{\int_{r_-}^{r_+} (h^2 |\partial_{r^*}f_i|^2 + V_{\text{eff}}^{(h)}(r) |f_i|^2 )\upd r^*}{||f_i||_{L^2}^2}
 \label{nth eigenvalue characterised}
\end{equation}
i.e.\ to find the $n$-th eigenvalue we calculate the maximum Rayleigh quotient among a set of $n$ mutually orthogonal functions in $H^1_0([r_-, r_+])$, and then minimize this among all such sets of $n$ functions. Note that the inner product $\langle \cdot, \cdots \rangle$ used above is the standard $L^2$ inner product, and so the $n$ functions are orthogonal in $L^2$.

We can characterise the eigenvalues associated with the problem $P^i_D$ as
\begin{equation}
 \lambda^i_n = \inf_{\substack{(f_1, f_2, \ldots, f_n), f_i \in H^1_0([r^i_-, r^i_+]) \\ ||f_i||_{L^2} \neq 0, \ \langle f_i, f_j \rangle = 0 \ \forall i\neq j}} \max_{i\leq n} \frac{\int_{r_-}^{r_+} (h^2 |\partial_{r^*}f_i|^2 + V_{\text{eff}}^{(h)}(r) |f_i|^2 )\upd r^*}{||f_i||_{L^2}^2}
\end{equation}
where $r^i_-$ and $r^i_+$ are the smallest and largest values of $r^*$ in the $i$-th interval. Similarly, we can characterise the eigenvalues associated with the problem $P^i_N$ as
\begin{equation}
\mu^i_n = \inf_{\substack{(f_1, f_2, \ldots, f_n), f_i \in H^1([r^i_-, r^i_+])\\ ||f_i||_{L^2}\neq 0, \ \langle f_i, f_j\rangle = 0 \ \forall i\neq j}} \max_{i\leq n}\frac{\int_{r_-}^{r_+} (h^2 |\partial_{r^*}f_i|^2 + V_{\text{eff}}^{(h)}(r) |f_i|^2 )\upd r^*}{||f_i||_{L^2}^2}
\end{equation}
where now we simply take the infemum over \emph{all} functions in $H^1([r^i_-, r^i_+])$.

Let us group the eigenvalues to the problems $P^i_D$, for \emph{all} $i$, into a single monotone non-decreasing sequence $\lambda_1, \lambda_2, \ldots$, and similarly let us group all the eigenvalues to the problems $P^i_N$, for all $i$, into a non-decreasing sequence $\mu_1, \mu_2, \ldots$. We claim that $\mu_n \leq \kappa_n \leq \lambda_n$. To show that $\kappa_n \leq \lambda_n$, let the eigenfunctions associated to the $\lambda_n$ be denoted $f_n$. Then, we extend the functions $f_n$ to functions $\tilde{f}_n$ on the whole of $[r_-, r_+]$ by setting $\tilde{f}_n = 0$ outside the range in which $f_n$ was originally defined\footnote{Note that functions defined in this way will not, in general, be $C^2$, even if the original function $f_n$ is. However, it is easy to check that one can always find a sequence of $C^2$ functions approaching $\tilde{f}_n$ in $H^1$, and so the functions $\tilde{f}_n$ \emph{are} in $H^1_0([r_-, r_+])$.}. Note that these functions $\tilde{f}_n$ are orthogonal to each other in $H^1$, either because they are different eigenfunctions which solve the same problem or because they are supported on different regions, and so we can immediately see from \ref{nth eigenvalue characterised} that $\kappa_n \leq \lambda_n $.

Next, to show that $\mu_n \leq \kappa_n$, consider the space $Y$ of functions defined by
\begin{equation}
\begin{split}
 Y = \Big\{ & f \in H^1([r_-, r_+]) \, \Big| \, ||f||_{L^2} \neq 0, \ f \text{ in the closure in }H^1\text{ of \emph{piecewise} } C^2 \text{ functions,} \\
 &\text{which are } C^2 \text{ in the intervals } [r_-^1, r_+^2], \ldots [r_-^k, r_+^k] \Big\}
\end{split}
\end{equation}
We define
\begin{equation}
 \tilde{\mu}_n = \inf_{\substack{(f_1, f_2, \ldots, f_n), \ f_i \in Y \\ \langle f_i, f_j \rangle \neq 0 \ \forall i \neq j}}\max_{i\leq n} \frac{\int_{r_-}^{r_+} (h^2 |\partial_{r^*}f_i|^2 + V_{\text{eff}}^{(h)}(r) |f_i|^2 )\upd r^*}{||f_i||_{L^2}^2}
\end{equation}
Evidently, we have that $H^1_0([r_-, r_+]) \subset Y$, and so it follows that $\tilde{\mu}_n \leq \kappa_n$. We only need to show that the eigenvalue $\tilde{\mu}_n$ is actually equal to the eigenvalue $\mu_n$ and the proof of the first part of the sublemma is complete. 

Recall that, from the variational characterisation of eigenvalues, $\tilde{\mu}_n$ is an eigenvalue associated to some eigenfunction $\tilde{f}_n \in Y$, while $\mu_n$ is an eigenvalue associated to an eigenfunction $f_n \in H^1([r_-^i, r_+^i])$ for some sub-interval $[r_-^i, r_+^i] \subset [r_-, r_+]$. The eigenfunctions $f_n$ solve the eigenvalue problem $P^i_N$, i.e.\ with Neumann boundary conditions imposed at $r_-^i, r_+^i$. As before, we can extend the functions $f_n$ to functions in $Y$ by setting $f_n = 0$ outside the interval $[r_-^i, r_+^i]$. Also from the variational characterisation of eigenvalues, we see that in each sub-interval $[r_-^j, r_+^j]$ the functions $\tilde{f}_n$ must either vanish, or solve the eigenvalue problem $P^j_N$.

Moreover, if the function $\tilde{f}_n$ is supported on more than one sub-interval, say on the intervals $[r_-^{j_1}, r_+^{j_2}]$ and $[r_-^{j_1}, r_+^{j_2}]$, then it must solve the eigenvalue problems $P^{j_1}_N$ and $P^{j_2}_N$ with the \emph{same} eigenvalue for each problem. If this were not the case, then it is easy to see that the Rayleigh quotient can be decreased by first writing the function $\tilde{f}_n = \tilde{f}_n^{(1)} + \tilde{f}_n^{(2)}$, where $\tilde{f}_n^{(1)}$ is supported only on the interval $[r_-^{j_1}, r_+^{j_2}]$ while $\tilde{f}_n^{(2)}$ is supported only on the interval $[r_-^{j_1}, r_+^{j_2}]$. Then, by considering linear combinations $A\tilde{f}_n^{(1)} + B\tilde{f}_n^{(2)}$ for constants $A$ and $B$, we can decrease the Rayleigh quotient, contradicting minimality. Consequently, $\tilde{f}_n^{(1)}$ solves the problem $P^{j_1}_N$ and $\tilde{f}_n^{(2)}$ solves the problem $P^{j_2}_N$ with the \emph{same} eigenvalue, and both $f_n^{(1)}$ and $f_n^{(2)}$ are functions in $Y$ which minimize the Rayleigh quotient characterizing the $n$-th eigenvalue.

From the above remarks it is clear that, when counted with multiplicity, $\mu_n = \tilde{\mu}_n$, and so the proof of the first part of sublemma \ref{sublemma on number of eigenvalues} is complete. Next, we need to show that increasing and decreasing $V_{\textnormal{eff}}$ leads to a corresponding increase or decrease in the eigenvalue $\kappa_n$. This follows trivially from the variational characterisation of the eigenvalues given in equation \ref{nth eigenvalue characterised}.

\vspace{1em}
The number of eigenvalues for the problems $P^i_D$ and $P^i_N$ can be calculated directly, since the corresponding problems can be solved exactly. Indeed, we obtain
\begin{equation}
\begin{split}
\sum_{i=1}^k N_{\leq E}(P^i_D) &= \sum_{i=1}^k \Big\lfloor \frac{\beta}{\pi h}\left(\sqrt{E-V^i_+}\right)\chi_{\{E\geq V^i_+\}} \Big\rfloor \\
&= \sum_{i=1}^k \left(\frac{\beta}{\pi h}\left(\sqrt{E-V^i_+}\right)\chi_{\{E\geq V^i_+\}} \right) + \mathcal{O}(k)
\end{split}
\end{equation}
Also,
\begin{equation}
\begin{split}
\sum_{i=1}^k N_{\leq E}(P^i_N) &= \sum_{i=1}^k \Big\lfloor \frac{\beta}{\pi h}\left(\sqrt{E-V^i_-}\right)\chi_{\{E\geq V^i_-\}} \Big\rfloor \\
&= \sum_{i=1}^k \left(\frac{\beta}{\pi h}\left(\sqrt{E-V^i_-}\right)\chi_{\{E\geq V^i_-\}} \right) + \mathcal{O}(k)
\end{split}
\end{equation}
Next let the number of cells tend to infinity as $h \rightarrow 0$ such that $k(h) = o(1/h)$, and the statement of the theorem follows.

\subsection{Energy Estimates}
We now state the energy estimate which will be used to capture the exponential decay of eigenfunctions of Schr\"odinger-type operators in the classically forbidden region:
\begin{lemma}
Let $r_1 > r_2$, $h>0$ and let $u$, $W$ and $\phi$ be smooth, real valued functions on $[r_1, r_2]$, with $u(r_1) = u(r_2) = 0$. Then
\begin{equation}
\begin{split}
&\int_{r_1}^{r_2}\left(\left| \frac{\upd}{\upd r^*}\left(e^{\phi/h} u\right) \right|^2 + h^{-2}\left(W - \left(\frac{\upd \phi}{\upd r^*}\right)^2\right) e^{2\phi/h}|u|^2\right)\upd r^* \\
&= \int_{r_1}^{r_2}\left(-\frac{\upd^2 u}{\upd r^{*2}} + h^{-2}W u\right) u e^{2\phi/h}\upd r^*
\end{split}
\end{equation}
\label{agmon1}
\end{lemma}
\emph{Proof}. Integrate by parts.

\subsection{Agmon Distance}

\begin{figure}[thb]
\centering
\includegraphics[width = 0.8\linewidth, keepaspectratio]{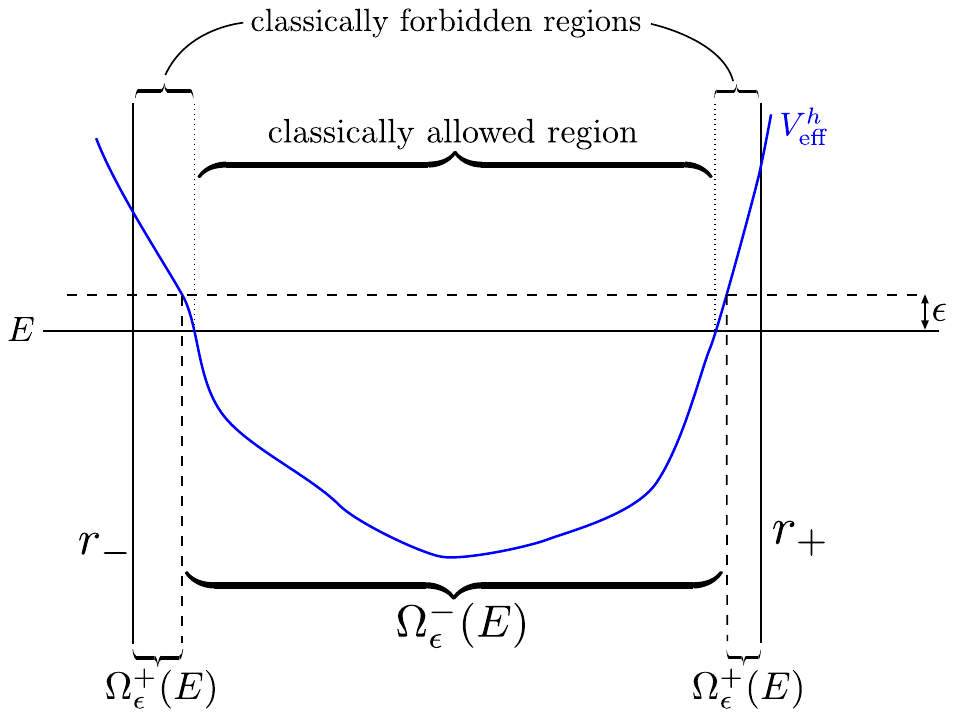}
\caption{The regions $\Omega_\epsilon^-(E)$, which is a slight enlargement of the classically allowed region, and $\Omega_\epsilon^+(E)$, which is slightly smaller than the classically forbidden region.}
\label{forbiddenregionfigure}
\end{figure}

Define the Agmon distance between two points:
\begin{equation}
d(r_1, r_2) = \left| \int_{r^*(r_1)}^{r^*(r_2)} \left(\sqrt{V_{\textnormal{eff}}^h(r^*) - E}\right)\chi_{\{ V_{\textnormal{eff}}^h \geq E\}} \upd r* \right|
\end{equation}
The Agmon distance satisfies
\begin{equation}
|\partial_{r^*} d(r^*, r_2)|^2 \leq \max\{V_{\textnormal{eff}}^h(r^*) - E, 0\}
\end{equation}
For any energy level $E$, we also define the distance to the classically allowed region by
\begin{equation}
d_E (r^*) = \inf _{r_1 \in \{E \geq V_{\textnormal{eff}}^h\}} d(r^*, r_1)
\end{equation}
Finally, we define the two (not necessarily connected) regions, illustrated in figure \ref{forbiddenregionfigure}:
\begin{equation}
\begin{split}
\Omega_\epsilon^+(E) &= \{ r^* : V_{\textnormal{eff}}^h(r^*) > E + \epsilon \} \cap [r_-,r_+] \\
\Omega_\epsilon^-(E) &= \{ r^* : V_{\textnormal{eff}}^h(r^*) \leq E + \epsilon\}
\end{split}
\end{equation}

\subsection{The Main Estimate for the Lower Bound}
The main estimate we need is an application of the energy estimate to solutions of (\ref{Schrodinger}), which quantifies the size of the solution in the forbidden region. This will be used to gain control over the errors incurred by cutting off the solution in this region.

\begin{lemma}
Let $u$ be a solution to (\ref{Schrodinger}), with Dirichlet boundary conditions imposed at $r_-$ and $r_+$ and eigenvalue $\kappa$ satisfying $|\kappa - E| \leq \delta$. For $\epsilon \in (0, 1)$, define
\begin{equation}
\phi_{E, \epsilon}(r*) = (1-\epsilon)d_{E}(r*)
\end{equation}
so that $phi$ is a slightly modified distance function to the region in which the effective potential is below $E$. Also define
\begin{equation}
a_E(\epsilon) = \sup_{\Omega_\epsilon^-(E)} d_{E} 
\label{define a}
\end{equation}
recalling that the region $\Omega_\epsilon^-(E)$ is a slight enlargement of the classically allowed region, in which the effective potential is bounded by $V^h_{\textnormal{eff}}(r*) \leq E + \epsilon$. Then for $\epsilon$ and $h$ sufficiently small, and for $\delta$ sufficiently small (depending on $\epsilon$ and $h$), $u$ satisfies
\begin{equation}
\int_{r^*(r_-)}^{r^*(r_+)} h^2 \left| \frac{\upd}{\upd r^*} e^{\phi_{E, \epsilon}/h} u\right|^2 \upd r^* + \frac{1}{2}\epsilon^2 \int_{\Omega_\epsilon^+}e^{2\phi_{E, \epsilon}/h}|u|^2 \upd r^* \leq \left(\kappa + \frac{1}{2}\epsilon\right)e^{2 a_E(\epsilon)/h} ||u||^2_{L^2(r_-, r_+)}
\label{main estimate}
\end{equation}
\label{main lemma}
\end{lemma}
\emph{Proof}. Apply lemma \ref{agmon1} to the region $r_- \leq r^* \leq r_+$, with $W(r*) = V^h_{\textnormal{eff}}(r*) - \kappa$ and $\phi = \phi_{E, \epsilon}$ to obtain
\begin{equation}
\begin{split}
\int_{r_-}^{r_+} & h^2 \Big| \frac{\upd}{\upd r^*} e^{\phi_{E,\epsilon}/h} u\Big|^2 \upd r^* + \int_{\Omega_\epsilon^+(E)} \left(V_{\textnormal{eff}}^h(r^*) - \kappa - \left|\frac{\upd\phi_{E,\epsilon}}{\upd r^*} \right|^2\right)e^{2\phi_{E,\epsilon}/h}|u|^2 \upd r^* \\
& = \int_{\Omega_\epsilon^-(E)} \left( \kappa - V_{\textnormal{eff}}^h(r^*) + \left|\frac{\upd\phi_{E,\epsilon}}{\upd r^*} \right|^2\right)e^{2\phi_{E,\epsilon}/h}|u|^2 \upd r^*
\end{split}
\end{equation}
In the regions $\Omega_\epsilon^+(E)$, we have
\begin{equation}
\begin{split}
V_{\textnormal{eff}}^h - \kappa - \Big|\frac{\upd\phi_{E,\epsilon}}{\upd r^*}\Big|^2 &\geq 
\epsilon^2 - \delta
\end{split}
\end{equation}
while in the region $\Omega_\epsilon^-(E)$, for $h$ sufficiently small we have
\begin{equation}
\begin{split}
&\int_{\Omega_\epsilon^-}\left(\kappa - V_{\textnormal{eff}}^h + \Big|\frac{\upd \phi_{E,\epsilon}}{\upd r^*}\Big|^2 \right)e^{2\phi_{E,\epsilon}/h}|u|^2\upd r^* 
\leq \left(\kappa + \epsilon(1-\epsilon)\right)e^{2a_E(\epsilon)/h}||u||^2_{L^2(r_-, r_+)}
\end{split}
\end{equation}
where we have used the fact that $V_{\textnormal{eff}}^h \rightarrow V_{\textnormal{null}} > 0$ uniformly in $h$, and so for sufficiently small $h$, $V_{\textnormal{eff}}^h > 0$. Now, taking $\epsilon \leq \frac{1}{2}$ and $\delta^2 \leq \frac{1}{2}\epsilon^2$ proves the lemma.

\subsection{Exponential Suppression in the Forbidden Region}
We now apply the main estimate in order to quantify the size of the solution in the classically forbidden region. We will obtain the result that the solution is exponentially suppressed in the angular momentum, and this provides the required accuracy for obtaining logarithmic decay as a lower bound.

We first focus on the second term on the left of equation (\ref{main estimate}), dropping the first term. We first choose some $E \in (V_{\textnormal{min}}, V_{\textnormal{min}}+c)$. Using lemma \ref{aux lemma 1}, for all sufficiently small $\delta, \delta'$, we can find some constant $c''$ such that $\phi_{E, \epsilon} \geq c''$ for all $r \in [r_-, r_- + \delta']$, and all $\kappa$ in $[E-\delta, E+\delta]$. Moreover, this constant can be chosen uniformly in $\epsilon$.

Now we note that $a(\epsilon)$, defined in (\ref{define a}), satisfies
\begin{equation}
a_E(\epsilon) \rightarrow 0 \textnormal{ \ as \ } \epsilon \rightarrow 0
\end{equation}
uniformly in $h$, for $h$ sufficiently small. Recall that $a_E(\epsilon)$ is the maximum Agmon distance to the classically allowed region (in which $V^h_{\textnormal{eff}} \leq E$), within the slightly enlarged classically allowed region $\Omega^-_\epsilon$, and in addition, recall that $V^h_{\textnormal{eff}} \rightarrow V_{\textnormal{null}}$ uniformly as $h \rightarrow 0$. In particular, this means that we can now \emph{fix} some $\epsilon > 0$, sufficiently small that $a(\epsilon) \leq c''/2$, for all sufficiently small $h$. Note that this $\epsilon$ can be chosen to depend only on $V_{\textnormal{null}}$. We then obtain a constant $C$, \emph{independent} of $h$, such that
\begin{equation}
\int_{r_-}^{r_- + \delta'} |u|^2 \upd r^* + \int_{r_+-\delta'}^{r_+} |u|^2 \upd r^* \leq C e^{-C/h}||u||^2_{L^2(r_-, r_+)}
\label{suppression1}
\end{equation}

Next we turn our attention to the first term on the left of equation (\ref{main estimate}). Dropping the second term now, we find
\begin{equation}
\int_{\Omega_\epsilon^+}h^2 e^{2\phi_{E,\epsilon}/h}\left(\frac{1}{h^2}\left(\frac{\upd \phi_{E,\epsilon}}{\upd r^*}\right)^2 |u|^2 + \frac{2}{h}\frac{\upd \phi_{E,\epsilon}}{\upd r^*} u \frac{\upd u}{\upd r^*} + \Big| \frac{\upd u}{\upd r^*} \Big|^2 \right) \upd r^* \leq e^{2a_E(\epsilon)/h}||u||^2_{L^2(r_-, r_+)}
\end{equation}
The first term on the left can be dropped since it is positive. For the second term, we use Cauchy-Schwarz together with the estimate just proved, (\ref{suppression1}). Then, the bound on $|\upd \phi_{\kappa, \epsilon}/\upd r^*|$, and the bounds on $\phi_{\kappa, \epsilon}$ and $a(\epsilon)$ just outlined leads to:
\begin{equation}
h^2 \int_{\Omega_\epsilon^+}\Big|\frac{\upd u}{\upd r^*}\Big|^2 \upd r* \leq Ce^{-C/h}||u||^2_{L^2(r_-, r_+)}
\end{equation}
and hence, for $\delta'$ sufficiently small depending on $\epsilon$
\begin{equation}
\int_{r_-}^{r_- + \delta'}\Big|\frac{\upd u}{\upd r^*}\Big|^2 \upd r* + \int_{r_+-\delta'}^{r_+}\Big|\frac{\upd u}{\upd r^*}\Big|^2 \upd r* \leq C h^{-2}e^{-C/h}||u||^2_{L^2(r_-, r_+)}
\end{equation}
Note that we can find some smaller constant $C'$ such that $h^{-2} e^{-C/h} \leq e^{-C'/h}$ for all sufficiently small $h$.

Combining these results with our version of Weyl's law (lemma \ref{Weyl law}), we obtain our desired result, showing that the size of our solution in the forbidden region is exponentially suppressed as $l\rightarrow \infty$, compared with the solution in the trapped region:

\begin{lemma}
Pick $E \in (V_{\textnormal{min}}, V_{\textnormal{min}} + c)$, where $c$ is as in lemma \ref{aux lemma 1}. Then, for all sufficiently small $\delta$ and $\delta'$, we can find a sequence of eigenfunctions $\{ u_\ell \}_{\ell\geq L}$ to $P_{DD}(r_-, r_+)$ with eigenvalues $\kappa_\ell$ such that, for $\ell$ sufficiently large, $\kappa_\ell \in [E - \delta, E + \delta]$, and moreover there exists a constant $C$ such that
\begin{equation}
\int_{r_-}^{r_- + \delta'}\left(\Big|\frac{\upd u_\ell}{\upd r^*}\Big|^2 + |u_\ell|^2\right)\upd r* + \int_{r_+-\delta'}^{r_+}\left(\Big|\frac{\upd u_\ell}{\upd r^*}\Big|^2 + |u_\ell|^2\right) \upd r* \leq C e^{-C\sqrt{\ell(\ell +1)}}||u_\ell||^2_{L^2(r_-, r_+)}
\label{suppression eqn}
\end{equation}
\label{suppression2}
\end{lemma}

\subsection{Constructing the Quasimodes}
Now that we have established the smallness of the solutions to the Dirichlet problem in the forbidden region, we can construct quasimodes by smoothly cutting off the solution in these regions. These quasimodes will then be approximate solutions to the wave equation, with correspondingly small errors. In this subsection we formalise these statements.

Let $\chi$ be a smooth function satisfying $\chi = 1$ on $[r_- + \delta', r_+-\delta']$ and $\chi = 0$ for $r^* \leq r_-$ and $r^* \geq r_+$. Then we define the \emph{quasimodes}:
\begin{equation}
\phi_\ell(t,r,\theta,\phi) = e^{i\omega_\ell t} \chi(r^*) r^{-1} u_\ell(r^*) S_{\ell 0}(\theta)
\label{quasimode}
\end{equation}
where $\omega_\ell = \ell(\ell + 1)$ and $\kappa_\ell$, $u_\ell$ are given by lemma \ref{suppression2}.

\begin{lemma}
There exists some $L$ such that, for $\ell > L$, the quasimodes $\phi_\ell$ are approximate solutions to the wave equation at all times, in the sense that they satisfy
\begin{equation}
||\Box_g \phi_\ell + F(r) \phi_\ell||_{H^k(\Sigma_t)} \leq C_k e^{-C_k \ell} ||\phi_\ell ||_{H^0 (\Sigma_0)}
\end{equation}
Moreover, the error is supported only in the strips $[r_-, r_- + \delta']$ and $[r_+ - \delta', r_+]$.
\label{quasimode error}
\end{lemma}
\emph{Proof}. By construction, $\Box_g \phi_\ell + F(r) \phi_\ell = 0$ outside of the strips stated in the lemma. Inside these strips, we can use our assumptions on the metric components to obtain the crude estimate
\begin{equation}
|\Box_g \phi_\ell + F(r) \phi_\ell| \leq C \left(\omega_\ell^2 |u_\ell| + \Big|\frac{\upd^2 u_\ell}{\upd r^{*2}}\Big| + \Big|\frac{\upd u_\ell}{\upd r^*}\Big| + h^{-2}|u_\ell| \right) S_{\ell 0}(\theta)
\end{equation}
If we integrate this inequality we can use equation (\ref{suppression eqn}) to bound all terms except for the one involving second derivatives, which can be controlled by using the equation satisfied by $u_\ell$, namely (\ref{Schrodinger}), together with our bounds on the effective potential.

For higher order derivatives, note that due to the form of the quasimodes (\ref{quasimode}) all time derivatives are trivially controlled by the $H^0$ norm of $u$. We can then commute the wave equation, and use the equation (\ref{Schrodinger}) every time two radial derivatives occur, and the equation satisfied by the $S_{\ell 0}$ every time two angular derivatives occur, to avoid terms with more than a single derivative appearing on the right hand side.

\subsection{Logarithmic Decay as a Lower Bound}
\label{log decay lower bound section}
Now that we have constructed the quasimodes, we show that we can construct a sequence of solutions to (\ref{wave equation 1}) which implies that \emph{no uniform decay estimate with a rate faster than logarithmic decay can hold}. By a ``uniform decay estimate'' we mean a statement of the form
\begin{equation}
\limsup_{t\rightarrow \infty} f(t) \int_{\Omega_1} e_1[\phi](t) r^2 \sin\theta \upd r\upd \theta\upd \phi \leq \int_{\Omega_2}e_2[\phi](0) r^2 \sin\theta\ \upd r\upd \theta\upd \phi
\end{equation}
Where this estimate is supposed to hold \emph{for all} $\phi$ solving (\ref{Schrodinger}) and arising from initial data in a suitable space. Here, $\Omega_1$ and $\Omega_2$ are fixed regions, with $\Omega_1$ bounded, and $e_1$ and $e_2$ are some chosen positive definite ``energy densities'' of the wave $\phi$. The function $f(t)$ captures the uniform rate of decay.

Note that ``\emph{uniform}'' is crucial here - it may well be the case that each individual solution in our sequence decays faster than logarithmically. Indeed, it may even be the case that all solutions with, say, smooth initial data decay faster than logarithmically. However, given any faster decay rate, it would be possible to find a smooth solution which decays \emph{slower} than this rate.

We now state our main theorem of this section:
\begin{theorem}
In a spherically symmetric spacetime admitting stably trapped null geodesics in the sense of section (\ref{stable trapping}), there exists some universal constant $C>0$ such that
\begin{equation}
\limsup_{\tau\rightarrow \infty} \, \sup_{\phi \neq 0} \ \log(2+\tau) \left( \frac{\mathcal{E}^T[\phi](\tau)\big|_{[r_-, r_+]}}{\mathcal{E}^T_2[\phi](0)} \right)^{\frac{1}{2}} > C
\end{equation}
where by $\mathcal{E}^T_2[\phi]\big|_{[r_-, r_+]}$ we mean the energy of the wave $\phi$ restricted to the region $\Sigma_\tau\cap[r_-, r_+]$, the surface $\Sigma_\tau$ is given by $t = \tau$, and the supremum is taken over all functions $\phi$ in the completion of the set of smooth, compactly supported functions with respect to the norm defined by the higher order energy, $E^T_2$.
\end{theorem}

\emph{Proof}. Denote by $P(\tau, s)(\overline{\phi}, \overline{\partial_t \phi})$ the solution, at time $\tau$, to the homogeneous problem
\begin{equation}
\begin{split}
\Box_g \phi + F(r)\phi &= 0 \\
\phi \big|_{\Sigma_s} &= \overline{\phi} \\
\partial_t \phi \big|_{\Sigma_s} &= \overline{\partial_t \phi}
\end{split}
\end{equation}
and donate by $\phi_f$ the solution to the inhomogeneous problem
\begin{equation}
\begin{split}
\Box_g \phi + F(r)\phi &= f \\
\phi \big|_{\Sigma_s} &= \overline{\phi} \\
\partial_t \phi \big|_{\Sigma_s} &= \overline{\partial_t \phi}
\end{split}
\end{equation}
then Duhamel's formula yields
\begin{equation}
\phi_f (\tau) = P(\tau,0)(\overline{\phi}, \overline{\partial_t \phi}) + \int_0^\tau P(\tau,s)(0, -a f) \upd s
\end{equation}

Next, note that the quasimodes $\phi_\ell$ provide (compactly supported) initial data for each $\ell$, and moreover, for sufficiently large $\ell$ they satisfy
\begin{equation}
\begin{split}
\Box_g \phi_\ell + F(r)\phi_\ell &= f_\ell \\
||f_\ell||_{H^k(\Sigma_t)} &\leq C_k e^{-C_k \ell}||\phi_\ell||_{L^2 (\Sigma_0)}
\end{split}
\end{equation}
For each $\ell$, let $\tilde{\phi}_\ell$ denote the solution to the \emph{homogeneous} problem with the same initial data as $\phi_\ell$. Then Duhamel's formula gives us
\begin{equation}
\left(\mathcal{E}^T[\phi_\ell - \tilde{\phi}_\ell](\tau)\big|_{[r_-, r_+])} \right)^{\frac{1}{2}} \leq t \sup_{s\in[0,\tau]} \left(\mathcal{E}^T[P(\tau,s)(0, -a f)](\tau)\big|_{[r_-, r_+])}\right)^{\frac{1}{2}}
\label{Duhamel estimate}
\end{equation}
We now recall our uniform boundedness statement (\ref{boundedness}). This means that the solution to the homogeneous problem is bounded by the initial data, i.e. there exists a some $C$ such that\footnote{In this case, we can take $C = 1$}, for all initial data,
\begin{equation}
\mathcal{E}^T[P(t, 0)(\overline{\phi}, \overline{\partial_t \phi})](\tau) \leq C \mathcal{E}^T[P(t,0)(\overline{\phi}, \overline{\partial_t \phi})](0)
\end{equation}
Returning to (\ref{Duhamel estimate}) and using uniform boundedness, we find
\begin{equation}
\begin{split}
\left(\mathcal{E}[\phi_\ell - \tilde{\phi}_\ell](\tau)\big|_{[r_-, r_+])} \right)^{\frac{1}{2}}&\leq t C ||a f_\ell||_{L^2(\Sigma_0)} \\
& \leq t C e^{-C \ell}||\phi_\ell||_{L^2(\Sigma_0 \cap [r_-, r_+])} \\
& \leq t C e^{-C \ell}\left(\mathcal{E}^T[\phi_\ell](0)\big|_{[r_-, r_+])} \right)^{\frac{1}{2}}
\end{split}
\end{equation}
where we have made use of lemma \ref{quasimode error}, and also used a Poincar\'e inequality and the compactness of the initial data to bound the $L^2$ norm in terms of the energy.

Since the norms of the quasimodes are constant in time, we can use the reverse triangle inequality at any time $t \leq \frac{e^{C\ell}}{2C}$ to obtain
\begin{equation}
\begin{split}
\left(\mathcal{E}^T[\tilde{\phi}_\ell](\tau)\big|_{[r_-, r_+]} \right)^{\frac{1}{2}} &\geq \frac{1}{2}\left(\mathcal{E}^T[\phi_{\ell}](\tau)\big|_{[r_-, r_+]} \right)^{\frac{1}{2}}\\
& \geq \frac{1}{2}\left(\mathcal{E}^T[\phi_{\ell}](0)\right)^{\frac{1}{2}} \\
& \geq \frac{C}{\ell}\left(\mathcal{E}^T_2[\phi_{\ell}](0)\right)^{\frac{1}{2}}
\end{split}
\label{triangle inequality}
\end{equation}
where we have used the spatial localisation of the initial data in the second line, and in the final line we have used frequency localisation and angular frequency localisation to exchange angular and time derivatives with factors of $\ell^{-1}$, and the wave equation to replace radial derivatives with angular and time derivatives. We have also used the fact that the initial data for the quasimodes and the solutions $\tilde{\phi}_\ell$ are the same.

Finally, setting $\tau = \frac{e^{C\ell}}{2C}$ we obtain a sequence of times $\tau_\ell$ and solutions $\tilde{\phi}_\ell$ satisfying
\begin{equation}
\left(\frac{\mathcal{E}^T[\tilde{\phi}_\ell](\tau_\ell)\big|_{[r_-, r_+]}}{\mathcal{E}^T_2[\tilde{\phi}_\ell](0)}\right)^{\frac{1}{2}} > C
\end{equation}
where the constant $C$ is independent of $\ell$. This proves the lemma.
\\[5pt]
Note that we can use frequency and angular frequency localisation, together with the equation satisfied by the radial part of the quasimodes, and the exponentially small error in the cut-off regions, to commute with an arbitrary (finite) number of derivatives, which then appear on the right hand side of (\ref{triangle inequality}) at the expense of additional factors of $\ell^{-1}$.

This point deserves some further discussion. Note that, in view of the (unstable) trapping at the photon sphere in Schwarzschild, one cannot prove a local energy decay statement of any kind \emph{without losing derivatives} (such statements are routinely made in the literature, and were finally proven in \cite{Sbierski2013a}). On the other hand, if we take one more angular derivative of the initial data (or even an $\epsilon$ of a derivative - see \cite{Blue2009} and also \cite{Dafermos2013}), then we \emph{can} prove a local decay statement - see \cite{Dafermos2009}. Thus the trapping at the photon sphere does not pose a serious problem as far as energy decay goes. One might worry that the situation could be similar in our case, however, in view of the above comments, we cannot expect to rescue a significantly faster local decay statement by taking additional derivatives of the initial data. In fact, it is easy to see that, if one wishes to bound the decay of the energy of the solution at a time $t$ in terms of up to $n$ derivatives of the initial data, then one cannot prove a uniform bound with a rate greater than $(\log{t})^{-(n-1)}$.

\subsection{Discussion}
We have succeeded in proving a link between stable trapping in spherically symmetric spacetimes, and slow decay of solutions to a linear wave equation on such a background. In particular, we have shown that one cannot obtain a uniform bound on the decay of solutions to the wave equation which is better than logarithmic.

We have already emphasised the importance of the word \emph{uniform} in the above statement. Indeed, the individual solutions we have constructed may well decay faster. Physically, we are considering quasibound states, and using the fact that the waves behave more and more like particles at high angular momentum to construct solutions which tunnel through the potential barrier at increasingly slow rates.

It is important to note that our solutions may not, in fact, decay at all. The only restriction we need on the metric is that it is spherically symmetric and regular in some region near a local minimum of the effective potential. As such, one could certainly construct spacetimes in which this local minimum is in fact a global minimum, and in such a spacetime one would expect the existence of \emph{normal modes}, which do not decay at all. However, if we consider asymptotically flat spacetimes, then the effective potential should approach zero at infinity, so, if the effective potential is positive, then in the absence of low-frequency obstructions we would expect some kind of decay. Our result says that we could not hope for anything better than logarithmic decay. In the subsequent part of this paper, we will indeed prove that logarithmic decay holds as an \emph{upper bound}, once some additional conditions on the metric and the function $F$ are imposed, which are sufficient to preclude low-frequency obstructions.

We now make some comparisons with \cite{Holzegel2014}. In \cite{Holzegel2013}, logarithmic decay was proved as an upper bound for Kerr-AdS, and so the results of \cite{Holzegel2014} showed that in fact this decay estimate is sharp. In the general case we consider, there is no hope of a similar result unless extra conditions are imposed, as they will be in the subsequent section. Indeed, in the asymptotically flat case there is some hope of a universal logarithmic decay estimate whenever uniform boundedness holds - see \cite{Burq1998}.

The results of this part of the paper are a fairly straightforward application of the methods of \cite{Holzegel2014}. The main difference is that we localise our quasimodes around some local minimum of the effective potential, whereas in \cite{Holzegel2014} the ``local'' minimum is located at infinity. As such, the quasimodes in Kerr-AdS are only semi-local, whereas in our case they actually have compact support\footnote{This is another instance of AdS space acting as a ``box''}. Note that, in our case, we also ``cut off'' the quasimodes in both directions, whereas in Kerr-AdS they are cut-off only at some inner boundary. In fact, if our metric is regular (in the chosen co-ordinates) down to $r=0$ we do not need to perform this - we could just localise our quasimodes in the region $r\in [0, r_+]$, and cut off at the outer boundary. We are also able to consider a large class of linear wave equations, rather than just the massive wave equation considered in \cite{Holzegel2014}. Finally, we provide a slightly more explicit link between the trapping of null geodesics and logarithmic decay. On the other hand, we have restricted to the spherically symmetric case, whereas the full Kerr-AdS family was considered in \cite{Holzegel2014}. Indeed, much of the technical work in \cite{Holzegel2014} actually involved performing various continuity arguments in order to extend the results to the rotating case. In some cases, a similar approach may work, but one would have to check each case individually.

The main restriction of this method is to spherically symmetric spacetimes. Outside the realm of spherical symmetry, a similar approach may be possible \emph{if the wave equation separates}, which is, unfortunately, still very restrictive. Even in such a case, one may have to use continuity arguments along similar lines to \cite{Holzegel2014}, which may not always work. It would be very nice to be able to make some statement about decay in a general spacetime exhibiting stable trapping.

\section{The Upper Bound}
\label{upper bound section}

The second mathematical section of this paper is concerned with proving that, under a few additional conditions, logarithmic decay holds as an upper bound on the uniform decay rate of solutions to (\ref{wave equation 1}). This means that, given initial data in an appropriate space (defined in section \ref{results section}), the solution to the wave equation (\ref{wave equation 1}) will actually decay, and at least logarithmically quickly. By the results of the previous section, this result is sharp.

There are two reasons for the additional conditions on the background spacetime and the function $F$ which we will need in this section. First, decay is inherently tied to \emph{global} properties of the spacetime, as can be seen, for example, by a comparison between the cases of Kerr (\cite{Dafermos2010a}, \cite{Dafermos2014}) and Kerr-AdS \cite{Holzegel2013}. Thus, we will now need the co-ordinate system to be global, and the metric to be asymptotically flat. This should be contrasted with the approach taken in the previous section, where the idea was to construct \emph{localised} solutions, which were independent of the global or asymptotic properties of the spacetime.

The second reason we need to impose additional restrictions is to avoid certain low-frequency obstructions to decay, which can lead to exponentially growing ``mode'' solutions. These kind of low-frequency phenomena cannot be understood from the point of view of geodesic flow or trapping, a high-frequency phenomenon. A variety of conditions could be imposed which prevent these phenomena from occurring, but rather than make any serious attempt to list these, we will instead impose a condition\footnote{In particular, we will bound the effective potential $V_{\text{eff}}^{(\ell)}$} which appears reasonable in the neutron-star case, and which is sufficient for the purpose, although it can certainly be weakened in various ways.

We note that, at the conclusion of this section, we will have proven logarithmic decay as both a lower and an upper bound on the uniform decay rate for linear waves. Stating things this way, however, is potentially misleading. Our results certainly do not mean that \emph{all} solutions decay logarithmically. Our upper bound means that, indeed, all solutions do decay at least logarithmically. On the other hand, many solutions could decay faster than this. Our lower bound only means that we can construct a sequence of solutions which decay at a rate arbitrarily close to logarithmically for an arbitrarily long time. These provide counterexamples to any faster decay rates which could be proposed.

\subsection{Statement of Results}
\label{results section}
We now summarize the main results of this section.

\begin{theorem}
\emph{(Integrated Local Energy Decay for Bounded Angular Frequencies)}

Let $\phi$ satisfy the wave equation (\ref{wave equation 1}) on a manifold with the metric (\ref{metric}). Let the co-ordinates $(t, r)$ be globally defined on the quotient of the manifold by the action of $SO(3)$, and furthermore let the metric satisfy the asymptotic flatness conditions (\ref{asymptotic flat}) and the function $F$ fall off fast enough, as specified in (\ref{asymptotic freedom}). Furthermore, let the effective potential for the wave equation, $V_{\text{eff}}^{(\ell)}$, as defined in (\ref{reduced1}), satisfy
\begin{equation}
V_- r^{-2} \leq \ell(\ell+1) V_{\text{eff}}^{(\ell)} \leq (\ell(\ell + 1) + 1)V_+ r^{-2}
\end{equation}
for some positive constants $V_-$ and $V_+$. Define the projection, $\phi^\flat$, of $\phi$ onto the angular frequency range $\ell(\ell + 1) \leq \lambda$. Then, for all sufficiently large $R$ and $\tau$ we have
\begin{equation}
\int_0^\tau \upd t \int\limits_{\Sigma_t \cap \{r\leq R\} }\left( \frac{1}{r^2}(\phi^\flat)^2 + J^T_\mu[\phi^\flat]n^\mu_{\Sigma_t}\right) \leq C_1 e^{C_2 \sqrt\lambda}\int_{\Sigma_0} J^T_\mu[\phi^\flat]n^\mu_{\Sigma_0}
\label{ILED equation}
\end{equation}
for some (uniform in time) constants $C_1$ and $C_2$, as long as the right hand side is finite.
\label{ILED theorem}
\end{theorem}

\begin{theorem}
\emph{(Energy Decay for Bounded Angular Frequencies)}

Under the assumptions of theorem \ref{ILED theorem}, if we also assume that the initial data is sufficiently localised so that
\begin{equation}
\int\limits_{\substack{u = u_0 \\ r\geq R}} \upd v \int_{\mathbb{S}^2} \upd\Omega \, r^2 (\partial_v \psi^\flat)^2 < \infty
\end{equation}
where $u$ and $v$ are null co-ordinates defined by $u = \frac{1}{2}(t-r^*)$, $v = \frac{1}{2}(t+r^*)$, and $u_0 := -\frac{1}{2}R^* = -\frac{1}{2}r^*(R)$, then, if $\Sigma_\tau$ is an appropriate foliation (see figure \ref{picture1} for details), the energy of $\phi^\flat$ decays as
\begin{equation}
\int_{\Sigma_\tau}\left(\frac{1}{r^2}(\phi^\flat)^2 + J^T_\mu[\phi^\flat]n^\mu_{\Sigma_\tau}\right) \leq C\tau^{-2}\left(\int_{\Sigma_0} J^T_\mu[\phi^\flat]n^\mu_{\Sigma_0} + \int\limits_{\substack{u = u_0 \\ r\geq R}} \upd v \int_{\mathbb{S}^2} \upd\Omega \, r^2 (\partial_v \psi^\flat)^2 \right)
\label{bounded energy decay equation}
\end{equation}
\label{bounded energy decay theorem}
\end{theorem}

\begin{theorem}
\emph{(Logarithmic Decay of Energy)}

Under the assumptions of theorem \ref{bounded energy decay theorem}, if we additionally assume that the data is sufficiently localised such that
\begin{equation}
\int_{\Sigma_0} \sum_{i=1}^3 J^T_\mu[\Omega_i \phi]n^\mu_{\Sigma_0} < \infty
\end{equation}
where the $\Omega_i$ are angular momentum operators, then the energy decays as
\begin{equation}
\begin{split}
\int_{\Sigma_\tau}\left(\frac{1}{r^2}\phi^2 + J^T_\mu[\phi]n^\mu_{\Sigma_\tau}\right) \leq \frac{C}{(\log\tau)^2} \Bigg( & \int_{\Sigma_0} \left(J^T_\mu[\phi]n^\mu_{\Sigma_0} + \sum_{i=1}^3 J^T_\mu[\Omega^i\phi]n^\mu_{\Sigma_0}\right) \\
& + \int\limits_{\substack{u = u_0 \\ r\geq R}} \upd v \int_{\mathbb{S}^2} \upd\Omega \, r^2 (\partial_v \psi)^2 \Bigg)
\end{split}
\label{log decay equation}
\end{equation}
\label{log decay theorem}
\end{theorem}

\begin{corollary}
\emph{(Pointwise Decay)}

Suppose the assumptions of theorem \ref{log decay theorem} hold, and additionally that they hold with $\phi$ replaced by $\sum_i \Omega_i \phi$. Then we obtain the following pointwise decay:
\begin{equation}
|\phi|^2 \big|_{\Sigma_\tau} \leq \frac{C}{(\log\tau)^2} \left(\int_{\Sigma_0} \sum_{|\alpha| \leq 2}J^T_\mu[\Omega^{(\alpha)}\phi]n^\mu_{\Sigma_0} + \int\limits_{\substack{u = u_0 \\ r\geq R}} \upd v \int_{\mathbb{S}^2} \upd\Omega \, \sum_{|\alpha|\leq 1} r^2 (\partial_v \Omega^{(\alpha)} \psi)^2 \right)
\end{equation}
where $(\alpha)$ is a multi index.
\label{pointwise decay corollary}

\end{corollary}

\subsection{Overview of the Proof}

There are a few ideas which go into our proof of the above theorems. We already have established the trivial non-degenerate energy bound (\ref{boundedness}), and so the next stage in our proof will naturally be to establish some kind of ``integrated local energy decay'' (ILED) statement. These kind of statements allow us to bound the energy, integrated over a \emph{spacetime} region, by the initial energy. Indeed, almost all of the work in this section goes into establishing such a statement. Once we have achieved this, and also established a ``weighted energy'' statement, the ``black box'' results of \cite{Dafermos2010b} can be used to establish decay.

We will actually only be able to achieve the above mentioned ILED statement for low frequencies\footnote{By ``low'' here, we actually mean bounded above, and that bound can be arbitrarily high, though not infinite.}. We will have to use a different argument in order to control the unbounded, high frequency range. In view of the stable trapping, and the links between trapping and energy decay (see \cite{Sbierski2013a}), we should expect that high frequencies are harder to control. Luckily, this argument, which exploits the largeness of the frequencies, will be sufficient to prove logarithmic decay.

We note here that it is often preferable to use a purely physical-space approach to proving decay, since such approaches are generally much simpler to use. In particular, given the relatively simple metrics which we consider, and the availability of physical-space approaches to wave equations on, for example, Schwarzschild \cite{Dafermos2010b}, this may appear to be possible. However, as indicated in section \ref{lower bound section}, it is \emph{high frequency} waves which form an obstruction to (fast) decay, and so it seems that we must make use of a frequency-space type approach.

The basic philosophy behind our proof is to make exclusive use of robust methods which have previously appeared in the literature. We have already mentioned the ``black box'' method of \cite{Dafermos2010b}, which allow us to conclude decay from a combination of energy boundedness, ILED and weighted energy results, and these are extremely robust, relying essentially on the asymptotic flatness of the metric and little else. In order to prove our ILED statement, we will make extensive use of another robust tool: exponentially weighted multipliers, which previously appeared in, for example \cite{Dafermos2010a} and \cite{Holzegel2013}. Although they are very robust, these estimates ``lose exponentially'' in the angular frequency, which is to say that the constants in the estimates depend exponentially on the angular frequency, in such a way that the estimates tell us nothing in the high frequency limit. This exponential loss will ultimately be responsible for the logarithmic decay.

It should be clear by the overview of the proof given above that we will have to split our solution into various frequency ranges. As mentioned in section \ref{cut off section}, in order to make sense of this we will first have to cut off our solution in time, and we will afterwards have to ensure that the errors this produces can be controlled.

\subsection{Time Cut-Off}
\label{cut off section}
We now return to discuss, in more detail, issues relating to the taking of the Fourier transform in time in section \ref{reduced section}. As has been already noted, we are not justified in making the \emph{a priori} assumption that a solution to the linear wave equation (\ref{wave equation 1}) is in $L^2$ in time. To ensure that we can indeed take the Fourier transform without running into difficulties, we will therefore cut-off the solution to the future and the past.

Define a smooth function $\xi(t)$ satisfying $\xi(t) = 0$, $t \leq 0$ and $\xi(t) = 1$, $t \geq 1$. Now, define $\xi_{\tau}(t) = \xi(t)\xi((\tau - t))$, and finally
\begin{equation}
\phi_{\hbox{\Rightscissors}}(t, \bm{x}) = \xi_{\tau}(t) \phi(t, \bm{x})
\end{equation}
Note that $\phi_{\hbox{\Rightscissors}}$ is now smooth and compactly supported in time, and so we can take the Fourier transform in time. Now, after applying the cut-off, $\phi_{\hbox{\Rightscissors}}$ no longer satisfies the wave equation, but instead satisfies
\begin{equation}
\Box \phi_{\hbox{\Rightscissors}} + F \phi_{\hbox{\Rightscissors}} = H
\end{equation}
where the function $H$ is supported in the two strips
\begin{equation}
\{ 0 \leq t \leq 1 \}\cup \{\tau - 1 \leq t \leq \tau \}
\end{equation}
and $H$ is given by
\begin{equation}
H = 2\nabla^\mu \xi_{\tau} \nabla_\mu \phi + (\Box \xi_{\tau})\phi
\end{equation}
and satisfies
\begin{equation}
\begin{split}
|\Box \xi_{\tau}| &\leq B \\
|\nabla^\mu \xi_{\tau} \nabla_\mu \phi| &\leq B\left( |\partial_t \phi|^2 + |\partial_{r^*}\phi|^2 + |\slashed{\nabla}\phi|^2 \right)
\end{split}
\label{H bounds}
\end{equation}
for some large $B$.

Note that, since $H$ is Schwarz, we can define $H_{m\ell}(r)$ by
\begin{equation}
ar H = \frac{1}{\sqrt{2\pi}}\int_\infty^\infty e^{-i\omega t}\sum_{m,\ell} P_{m \ell}(\cos \theta)e^{im\phi} H_{m\ell}(r)
\end{equation}
then we can correct the formula for the reduced equation (\ref{reduced1}) to
\begin{equation}
\partial^2_{r^*} u_{m\ell} + (\omega^2 - h^{-2} V_{\text{eff}}^{(h)} ) u_{m\ell} = H_{m\ell}
\label{reduced cut}
\end{equation}

We remark once again that, in order to prove our lower bound on the decay rate, we actually \emph{constructed} quasimodes, which (by construction) are approximate solutions to the unmodified reduced equation, (\ref{reduced1}). This avoids all the issues involving a time cut-off. In contrast, it cannot be avoided now that we are attempting to prove an upper bound for decay, since we are making statements about \emph{all} solutions to the wave equation (\ref{wave equation 1}), and we do not know whether these are in fact in $L^2$ in time.

\subsection{Integrated Local Energy Decay for Bounded Angular Frequencies}
Establishing a proof of ``integrated local energy decay'' is the largest and most difficult step necessary in order to eventually prove theorem \ref{log decay theorem}. This is a statement of the form
\begin{equation}
\int_{\tau = 0}^{\tau_1} \upd\tau \int\limits_{\Sigma_{\tau} \cap \{r \leq R \}} J^T[\phi] \lesssim \mathcal{E}^T[\phi](0)
\label{ILED schematic}
\end{equation}
Here, $R$ is some large but otherwise arbitrary constant. Actually, the presence of trapping prevents us from obtaining such a statement, as was proved in \cite{Sbierski2013a}. With this obstruction in mind, a common approach is to commute the initial energy a sufficient number of times with the Killing vector fields, obtaining a statement of the form
\begin{equation}
\int_{\tau = 0}^{\tau_1} \upd\tau \int\limits_{\Sigma_{\tau} \cap \{r \leq R \}} J^T[\phi] \lesssim \sum_{\Gamma \in S} \sum_{|\alpha| \leq n} \mathcal{E}^T[\Gamma^{(\alpha)} \phi]
\end{equation}
where the $S$ is an appropriate set of Killing vector fields, and $I$ is a multi-index, and $n$ is chosen sufficiently large. In fact, in our case we cannot achieve such a result either, which may be seen from our construction of the lower bound.

We will have to take a different approach in this paper. We \emph{will} prove a statement of the form (\ref{ILED schematic}), but not for the whole solution. Instead, only the projection of the solution onto low angular frequencies will satisfy an equation of the form (\ref{ILED schematic}). We will then rely on a different argument, which does not require an ILED-type statement, to handle the high frequencies.

Note that, throughout this section, we will fix some maximum angular frequency, $\ell(\ell + 1)\leq \lambda$, and keep track of the dependence of the various constants on $\lambda$. We will use robust methods, but these come at the price of very bad (exponential) dependence on $\lambda$ - nevertheless, this will be sufficient to prove logarithmic decay. We also note that we will only be concerned with proving a \emph{finite} ILED type statement, i.e.\ a statement of the form (\ref{ILED schematic}) for $R < \infty$. We will not need to establish the stronger result, in which we can send $R$ to infinity at the price of various weights in $r$, since we will instead use the ``new method'' of \cite{Dafermos2010b} to handle the asymptotic region.

We will use three different methods to get to our ILED result. First, we will produce an energy current which will obtain the required result for sufficiently large $r$, and which does not rely on the frequency decomposition. Next, we will find a frequency localised current which works for sufficiently small (time) frequencies, and bounded angular frequencies. Finally, we will use the exponentially weighted multipliers to produce a current for all time frequencies which are bounded away from both zero and infinity, and for bounded angular frequencies.

\subsubsection[A Large r Current]{A Large $r$ Current}

We have already defined the energy current associated to a vector field $X$:
\begin{equation}
J^X_\mu[\phi] = Q_{\mu\nu}[\phi] X^\mu
\end{equation}
and we now define the modified energy current
\begin{equation}
J^{X, w}_\mu[\phi] = Q_{\mu\nu}[\phi] X^\mu + \frac{1}{4}w \partial_\mu(\phi^2) - \frac{1}{4}(\partial_\mu w)\phi^2
\end{equation}
for some function $w$ to be specified. If $\phi$ satisfies the wave equation (\ref{wave equation 1}), then the modified energy current satisfies
\begin{equation}
K^{X, w}[\phi] := \mathop{\mbox{div}} J^{X, w}[\phi] = K^X[\phi] + \frac{1}{2} w \left( \partial^\mu \phi \partial_\mu \phi - F\phi^2 \right) - \frac{1}{4}(\Box w)\phi^2
\end{equation}
where 
\begin{equation}
K^X[\phi] := \mathop{\mbox{div}} J^X[\phi] = \frac{1}{2} X(F) + \frac{1}{2}Q^{\mu\nu}[\phi] (\mathcal{L}_X g)_{\mu\nu}
\end{equation}

Taking now $X = f(r^*) \partial_{r^*} = f a^{\frac{1}{2}}E^1$, we compute
\begin{equation}
\begin{split}
K^X &= \left( \frac{1}{2}f' + f a^{\frac{1}{2}}b^{-\frac{1}{2}}r^{-1}\right) (E^0(\phi))^2 + \left(\frac{1}{2}f' - f a^{\frac{1}{2}}b^{-\frac{1}{2}} r^{-1}\right) (E^1(\phi))^2 \\
&\phantom{=} + \left(-\frac{1}{2}f' - \frac{1}{2}f a^{-\frac{1}{2}}b^{-\frac{1}{2}}(\partial_r a)\right)\left( (E^2(\phi))^2 + (E^3(\phi))^2 \right) \\
&\phantom{=} + \left(\left(\frac{1}{2}f' + \frac{1}{2} f a^{-\frac{1}{2}}b^{-\frac{1}{2}}(\partial_r a) + f a^{\frac{1}{2}}b^{-\frac{1}{2}} r^{-1} \right)F + \frac{1}{2}f a^{\frac{1}{2}}b^{-\frac{1}{2}}(\partial_r F)\right)\phi^2
\end{split}
\end{equation}

Now, if we take $w = f' + 2fa^{\frac{1}{2}}b^{-\frac{1}{2}}r^{-1} - \delta a^{\frac{1}{2}}b^{-\frac{1}{2}}r^{-1-\delta} f$ for some $\delta > 0$, we find
\begin{equation}
\begin{split}
K^{X, w} &= \frac{1}{2}\delta a^{\frac{1}{2}}b^{-\frac{1}{2}}r^{-1-\delta} f (E^0(\phi))^2 + \left(f' - \frac{1}{2}\delta a^{\frac{1}{2}}b^{-\frac{1}{2}}r^{-1-\delta}f\right)(E^1(\phi))^2 \\
&\phantom{=} + \left(-\frac{1}{2}a^{-\frac{1}{2}}b^{-\frac{1}{2}}r^2 \partial_r(ar^{-2}) f - \frac{1}{2}\delta a^{\frac{1}{2}}b^{-\frac{1}{2}}r^{-1-\delta}f\right)|\slashed{\nabla}\phi|^2 \\
&\phantom{=} + \left(\left(-\frac{1}{2}a^{-\frac{1}{2}}b^{-\frac{1}{2}}(\partial_r a) + \frac{1}{2}\delta a^{\frac{1}{2}}b^{-\frac{1}{2}} r^{-1-\delta}\right)F + \frac{1}{2}f a^{\frac{1}{2}} b^{-\frac{1}{2}}F' - \frac{1}{4}\Box w \right)\phi^2
\label{bulk integral large r}
\end{split}
\end{equation}

Finally, we choose $f = (1-r^{-\delta})\chi_{R_0}$, where $\chi_{R_0}(r)$ is a smooth function which vanishes for $r \leq R_0-1$, and $\chi_{R_0} = 1$ for $r \geq R_0$. Now, with our asymptotic flatness conditions, (\ref{asymptotic flat}) and (\ref{asymptotic freedom}), it is easy to see that, choosing $R_0$ sufficiently large, we have
\begin{equation}
K^{X, w} \geq b(\delta)\left( r^{-1-\delta}(E^0(\phi))^2 + r^{-1-\delta}(E^1(\phi))^2 + r^{-1}|\slashed{\nabla}\phi|^2 + r^{-3-\delta}\phi^2\right)
\end{equation}
Hence, making use of the energy identity for the current $J^{X, w}[\phi]$, together with our energy boundedness statement (\ref{boundedness}) we can obtain the following:

\begin{lemma}
Let $\phi$ be a solution to the wave equation (\ref{wave equation 1}), decaying appropriately at infinity, with the metric function $a$ and $b$ and the function $F$ obeying the asymptotics (\ref{asymptotic flat}) and (\ref{asymptotic freedom}) respectively. Then there exists some constant $R_0$ such that, for all $\tau \geq 0$,
\begin{equation}
\begin{split}
\int_0^\tau \upd t &\int\limits_{\Sigma_t \cap \{ r\geq R_0 \}} \left( r^{-1-\delta}(E^0(\phi))^2 + r^{-1-\delta}(E^1(\phi))^2 + r^{-1}|\slashed{\nabla}\phi|^2 + r^{-3-\delta}\phi^2\right) \\
& \leq B(\delta)\mathcal{E}^T[\phi](0) + B(\delta)\int_0^\tau \upd t \int\limits_{\Sigma_t \cap \{R_0 - 1 \leq r \leq R_0 \}} \left((E^0(\phi))^2 + (E^1(\phi))^2 +|\slashed{\nabla}\phi|^2 +\phi^2\right)
\end{split}
\label{large r equation}
\end{equation}
\label{large r lemma}
\end{lemma}

We will make repeated use of this ``large $r$'' current later on. Since we have not used the frequency decomposition, we can use this current for all frequency ranges. Note, however, that the spacetime integral on the right hand side will, at some point, need to be absorbed by some other term before we can prove an ILED statement. In the subsequent sections we will construct frequency-localised currents, which have a good sign for $r$ arbitrarily large, but then a bad sign in some intermediate region. We will use these to absorb the spacetime term in (\ref{large r equation}), and simultaneously absorb their own bad spacetime integral terms (which occur at much larger values of $r$ than $R_0$) by using the large $r$ current.

\subsubsection{Frequency Localised Current Templates}

We consider a function $u$ satisfying
\begin{equation}
\partial^2_{r^*} u + (\omega^2 - V_\ell(r^*)) u = H(r^*)
\end{equation}
The current templates we will employ in the different frequency ranges have already appeared, in \cite{Dafermos2010a} and \cite{Holzegel2013} for example, but we state them here for completeness. They are:
\begin{equation}
\begin{split}
Q^f_0 &= f\left( |u'|^2 + (\omega^2 - V_\ell)|u|^2\right) + f'\mathop{\mbox{Re}}(u'\bar{u}) - \frac{1}{2} f''|u|^2 \\
Q^h_1 &= h\mathop{\mbox{Re}}(u'\bar{u}) - \frac{1}{2}h'|u|^2 \\
Q^g_2 &= g\left( |u'|^2 + (\omega^2 - V_\ell)|u|^2\right)
\end{split}
\end{equation}
and they satisfy
\begin{equation}
\begin{split}
(Q_0^f)' &= 2f'|u'|^2 - (f V_\ell' + \frac{1}{2}f^{\prime\prime\prime})|u|^2 + 2f\mathop{\mbox{Re}}(\bar{u}' H) + f'\mathop{\mbox{Re}}(\bar{u}H) \\
(Q_1^h)' &= h\left(|u'|^2 - \omega^2|u|^2 + V_\ell|u|^2 + \mathop{\mbox{Re}}(\bar{u}H)\right) - \frac{1}{2}h''|u|^2 \\
(Q_2^g)' &= g'\left( |u'|^2 + (\omega^2 - V_\ell)|u|^2\right) - g V_\ell'|u|^2 + 2g\mathop{\mbox{Re}}(\bar{u}'H)
\end{split}
\end{equation}
We will also make use of the quantity $Q^{g,h} := Q_1^h + Q_2^g$.

Inspired by \cite{Dafermos2010a}, we will now divide the different frequencies into two ranges: one, in which the (time) frequencies are bounded away from both zero and infinity (the \emph{bounded frequency} range), and in which we can use the robust exponentially weighted multipliers, and the other, in which the frequency is almost zero (the \emph{almost stationary} range).

\subsubsection{A Current for Almost Stationary Frequencies}
In this section, we seek a current for the frequency range $0 \leq \omega \leq \sqrt{\lambda}\omega_0$, $l(l+1)\leq\lambda$, where $\omega_0 >0$ can be taken as small as we like. As mentioned above, we will obtain a current with a good sign for all $r \leq R_1$, where $R_1$ can be chosen as large as we like. In fact, for this current, we will obtain a good sign everywhere, but it is only in the region $r \leq R_1$ that it will be coercive - we will need to add some of our ``large $r$'' current in order to eventually prove an estimate involving terms comparable to the energy.

First, note that
\begin{equation}
(Q^{g,h})' = (h + g')|u'|^2 + h(V_\ell - \omega^2)|u|^2 - (gV_\ell)'|u|^2 - \frac{1}{2}h''|u|^2 + h\mathop{\mbox{Re}}(\bar{u}H) + 2g\mathop{\mbox{Re}}(\bar{u}'H)
\end{equation}
Let us choose an $R_1$ such that, for all $r\leq R_1$, for all $\omega \leq \sqrt{\lambda}\omega_0$ and for all $\ell(\ell + 1)\leq \lambda$ we have
\begin{equation}
V_\ell - \omega^2 \geq c > 0
\end{equation}
Now, in $r \leq R_1$, we choose
\begin{equation}
\begin{split}
h &= R_1^{-2} \frac{r^*}{\rho + r^*} \\
\Rightarrow -\frac{1}{2}h'' &= R_1^{-2} \frac{\rho}{(\rho + r^*)^3}
\end{split}
\end{equation}
for some large $\rho$ to be fixed later, and
\begin{equation}
\begin{split}
g' + R_1^{-2}\frac{r^*}{\rho + r^*} \geq c > 0 \\
|(gV_\ell)'| \leq R_1^{-2} \frac{\rho}{(\rho + r^*)^3}
\end{split}
\end{equation}
so that, in $r \leq R_1$ we have
\begin{equation}
(Q^{g,h})' \geq c(|u'|^2 + |u|^2) + h\mathop{\mbox{Re}}(\bar{u}H) + 2g\mathop{\mbox{Re}}(\bar{u}'H)
\end{equation}

Next, for any $\epsilon_1 = \epsilon_1(\omega_0) > 0$ sufficiently small such that $-V_\ell' > 0$ for all $\lambda$ and for all $r \geq R_1 e^{\epsilon_1^{-1}}$, we choose, in the region 
$R_1 < r \leq R_1 e^{\epsilon_1^{-1}}$
\begin{equation}
\begin{split}
0 &\leq h \leq R_1^{-2} \\
|h'| &\leq 4 R_1^{-2} \epsilon r^{-1} \\
|h''| &\leq 4 R_1^{-2} \epsilon r^{-2} \\
h(R_1 e^{\epsilon_1^{-1}}) &= 0
\end{split}
\end{equation}
and
\begin{equation}
\begin{split}
g' &\geq V_- R_1^{-2} \\
-(gV)' &\geq V_- R_1^{-2} r^{-2} \\
g(R_1 e^{\epsilon_1^{-1}}) &= 1
\end{split}
\end{equation}
so that, in $R_1 \leq r\leq R_1 e^{\epsilon_1^{-1}}$ we have
\begin{equation}
(Q^{g,h})' \geq V_-R_1^{-2}|u'|^2 + \left(V_-r^{-2} -2\epsilon r^{-2} - \omega_0^2\right)R_1^{-2}|u|^2 + h\mathop{\mbox{Re}}(\bar{u}H) + 2g\mathop{\mbox{Re}}(\bar{u}'H)
\end{equation}
and we can ensure the positivity of the coefficient of $|u|^2$ in the region in question if we choose both $\epsilon_1$ and $\omega_0$ sufficiently small in comparison to $V_-$, and we also ensure
\begin{equation}
\epsilon > -\left(\frac{1}{2} \log(\omega_0^2 V_-^{-1} R_1^2)\right)^{-1}
\end{equation}

Finally, in the region $r\geq R_1 e^{\epsilon_1^{-1}}$ we pick $g = 1$, $h = 0$. Then
\begin{equation}
(Q^{g, h})' \geq r^{-3}V_+ |u|^2 + 2\mathop{\mbox{Re}}(\bar{u}'H)
\end{equation}
If we choose $\omega_0$ and $\epsilon_1$ small enough, and $R_1$ large enough, then we can indeed find smooth functions $g, h$ obeying the required inequalities.

Putting this all together, and integrating up to $R_+$, we obtain the following:
\begin{equation}
\begin{split}
c(\omega_0, \lambda, R)\int_0^{R_1} \left(|u'|^2 + (\omega^2 + \ell(\ell + 1) + 1)|u|^2\right) \leq& \int_0^{R_+} \left(h\mathop{\mbox{Re}}(\bar{u}H) + 2g\mathop{\mbox{Re}}(\bar{u}'H)\right) \\
& + Q^g_2({R_+}) - Q^{g,h}(0)
\end{split}
\label{almost stationary estimate}
\end{equation}
As already noted, we will only be interested in taking ${R_+}$ large and finite, and so we do not need to worry about $r$ weights in (\ref{almost stationary estimate}). Also, note the bad dependence on $\lambda$ of $b$ as $\lambda \rightarrow \infty$, $b\sim \lambda^{-1}$, which occurs as a result of our artificially introducing the term $\ell(\ell + 1)$, which we will use to bound the angular derivatives. However, for the ``high'' bounded frequencies, we will obtain even worse (exponential) control of the dependence of the corresponding constant on $\lambda$, and so this need not concern us.

\subsubsection{A Current for Intermediate Frequencies}
In this subsection, we will prove an estimate for all time frequencies bounded away from zero\footnote{Although we will take the lower bound for $\omega$ to be the same $\sqrt{\lambda}\omega_0$ as in the previous section, our estimates in this section would actually work given any lower bound strictly greater than zero.}, i.e.\ for $\omega \geq \sqrt{\lambda}\omega_0$, but still in our angular frequency range $\ell(\ell + 1)\leq \lambda$. Unlike in the ``almost stationary'' case, we will only be able to obtain a good sign for $r \leq R_2$, where $R_2$ can be chosen arbitrarily large. However, the negative terms in $r \geq R_2$ can be controlled by our large $r$ current.

First, pick any large $R_2$ and any $\epsilon_2 > 0$, and decompose $V_\ell = V_{\text{flat}} + V_{\text{junk}}$, where $V_{\text{flat}} \geq 0$, $V_{\text{junk}} \geq 0$ satisfy
\begin{equation}
\begin{split}
V_{\text{flat}} &= 0 \text{\quad for \quad} r\geq \epsilon_2^{-1}R_2 \\
V_{\text{junk}} &= 0 \text{\quad for \quad} r\leq R_2 \\
V_{\text{junk}}' &< 0 \text{\quad for \quad} r \geq \epsilon_2^{-1} R_2
\end{split}
\end{equation}
We also choose some $V_{\text{ind}}$, independent of $\ell$ (but still dependent on $\lambda$) such that, for all $\ell(\ell + 1) \leq \lambda$:
\begin{equation}
\begin{split}
V_-\lambda r^{-2} \leq V_{\text{flat}}\leq V_{\text{ind}}\leq V_+ (\lambda+1) r^{-2} &\text{\quad for \quad} r \leq R_2 \\
V_{\text{flat}} \leq V_{\text{ind}}\leq V_+ (\lambda+1) r^{-2} &\text{\quad for \quad} r \leq \epsilon_2^{-1}R_2 \\
V_{\text{ind}} = 0 &\text{\quad for \quad} r \geq \epsilon_2^{-1} R_2
\end{split}
\end{equation}

Now, suppose we have some $g \geq 0$ with $g' \geq 0$. Then
\begin{equation}
\begin{split}
\int_0^{R_+}(g V_{\text{flat}})'|u|^2 \upd r^* &= -\int_0^{R_+} 2g V_{\text{flat}}\mathop{\mbox{Re}}(\bar{u}u')\upd r^* + (g V_{\text{flat}}|u|^2)\big|_{r={R_+}} - (g V_{\text{flat}}|u|^2)\big|_{r=0}\\
&\leq \frac{1}{2}\int_0^{R_+} g'|u'|^2 \upd r^* + 2\int_0^{R_+} (g^2 V_{\text{flat}}^2 g')|u|^2 \upd r^* + (gV_{\text{flat}}|u|^2)\big|_{r={R_+}} \\
\end{split}
\end{equation}
we now choose
\begin{equation}
\begin{split}
g(r^*) &= \exp\left(-2\lambda^{-1/2}\omega_0^{-1}\int_{r^*}^\infty V_{\text{ind}}(\tilde{r}^*)\upd\tilde{r}^*\right) \\
\Rightarrow g' &= 2\lambda^{-1/2}\omega_0^{-1} V_{\text{ind}} g
\end{split}
\end{equation}
and so we obtain
\begin{equation}
\int_0^{R_+} (g V_{\text{flat}})'|u|^2 \upd r^* \leq \frac{1}{2}\int_0^{R_+}\left( g'|u'|^2 + \lambda\omega_0^2 g'|u|^2 \right)\upd r^* + (g V_{\text{ind}}|u|^2)\big|_{r={R_+}}
\end{equation}

Returning to our microlocal currents, we find that, with this choice of $g$,
\begin{equation}
\begin{split}
Q_2^g({R_+}) - Q_2^g(0) &= \int_0^{R_+} \Big( g'|u'|^2 + \omega^2 g'|u|^2 - (g V_{\text{flat}})'|u|^2 - g'V_{\text{junk}}|u|^2 - gV_{\text{junk}}'|u|^2 \\
& \phantom{\int_0^{R_+} \Big( } + 2g\mathop{\mbox{Re}}(\bar{u}'H) \Big)\upd r^* \\
&\geq \int_0^{R_+} \left(\frac{1}{2} g'|u'|^2 + \frac{1}{2}\omega^2 g'|u|^2 - (g'V_{\text{junk}} + gV_{\text{junk}}')|u|^2 + 2g\mathop{\mbox{Re}}(\bar{u}'H)\right)\upd r^* \\
&\phantom{\geq} - \left( gV_{\text{ind}}|u|^2\right)\big|_{r={R_+}}
\end{split}
\end{equation}
Now, we note that
\begin{equation}
\begin{split}
g' \geq V_- \lambda r^{-2}e^{-2b\omega_0^{-1}r^{-1}\sqrt{\lambda}} &\text{\quad in \quad} r\leq R_2 \\
0 \leq g' \leq V_+(\lambda+1)(\lambda)^{-1/2}\omega_0^{-1}r^{-2} &\text{\quad in \quad} R_2\leq r \leq \epsilon_2^{-1}R_2 \\
g'=0 \text{\, ,\, } g = 1 &\text{\quad in \quad} r\geq \epsilon_2^{-1}R_2
\end{split}
\end{equation}
We can also arrange for the following bounds to hold:
\begin{equation}
\begin{split}
0 \leq V_{\text{junk}} \leq V_+(\lambda + 1) r^{-2} \text{\, , \,} |V_{\text{junk}}'| \leq V_+ (\lambda + 1)\epsilon r^{-2} &\text{\quad in \quad} R_2 \leq r \leq \epsilon_2^{-1}R_2 \\
-V_{\text{junk}}' \geq V_- \lambda r^{-3} & \text{\quad in \quad} r \geq \epsilon_2^{-1}R_2
\end{split}
\end{equation}
Thus, after integrating and using $\omega^2 \geq \lambda\omega_0^2$ we obtain
\begin{equation}
\begin{split}
Q_2^g({R_+}) - Q_2^g(0) &\geq \int_0^{R_2} \left( V_- \lambda r^{-2} e^{-2V_-\sqrt{\lambda}\omega_0^{-1}r^{-1}}\left(\frac{1}{2}|u'|^2 + \frac{1}{4}\omega^2|u|^2 + \frac{1}{4}\lambda\omega_0^2|u|^2 \right)\right)\upd r^* \\
&\phantom{\geq} - \int_{R_2}^{\epsilon_2^{-1}}\left(V_+^2\omega_0^{-1}(\lambda)^{-1/2}(\lambda+1)^2 r^{-4} - V_+(\lambda + 1)\epsilon_2 r^{-2}\right)|u|^2\upd r^* \\
&\phantom{\geq} + \int_{\epsilon_2^{-1}R_2}^{R_+} V_- \lambda r^{-3}|u|^2 \upd r^* + \int_0^{R_+} + 2g\mathop{\mbox{Re}}(\bar{u}'H)\upd r^*
\end{split}
\end{equation}
which is almost a bound of the required type, except for the rather strange degeneration at $r = 0$. To fix this, we simply add another current, $Q_2^{\hat{g}}$, where
\begin{equation}
\begin{split}
\hat{g} = \hat{\epsilon}\frac{r^*}{\hat{\rho} + r^*} &\text{\quad for \quad} r\leq R_2 \\
|\hat{g}| \leq \hat{\epsilon} \text{\, , \,} |\hat{g}'| \leq 2 \epsilon_2 \hat{\epsilon} &\text{\quad for \quad} R_2 \leq r \leq \epsilon_2^{-1} R_2 \\
\hat{g} = 0 &\text{\quad for \quad} r\geq \epsilon_2^{-1} R_2
\label{preliminary bounded current}
\end{split}
\end{equation}
for sufficiently small $\hat{\epsilon}$ and sufficiently large $\hat{\rho}$. Then, for $\ell \neq 0$ we use the fact that
\begin{equation}
V_\ell \sim \ell(\ell + 1)r^{-2}
\end{equation}
to show that $(Q_2^{\hat{g}})'$ has a good sign and does not degenerate near $r = 0$. Away from $r = 0$, if we have chosen $\hat{\rho}$ and $\hat{\epsilon}$ appropriately, then all negative terms can be absorbed by (\ref{preliminary bounded current}).

Finally, we have to deal with the case $\ell = 0$, or spherical symmetry. In this case, we can actually avoid the time frequency localisation altogether, as it is easy to construct a current (modelled, say, on the Schwarzschild current of \cite{Dafermos2013}) for the spherically symmetric mode. Indeed, the only reason we cannot perform this construction for the entire frequency range is the presence of the term $\partial_r(a r^{-2})$, which appears, for example, in (\ref{bulk integral large r}), and which has a different sign depending upon whether the effective potential has a local minimum or a local maximum. However, note that in (\ref{bulk integral large r}), this term appears multiplying \emph{angular} derivatives, which vanish in the spherically symmetric case.

Noting that $Q_2^g(0) = 0$, we are finally led to the following estimate:
\begin{equation}
\begin{split}
& \int_0^{R_2}\left(|u'|^2 + (\omega^2 + \ell(\ell + 1) + 1)|u|^2\right)\upd r^* \\
\leq & R_2^2 C(\omega_0, \lambda)\left(\int_{R_2}^{\epsilon_2^{-1}R_2}\left(\epsilon r^{-2} + r^{-4}\right)|u|^2 \upd r^* + \int_{\epsilon_2^{-1}R_2}^{{R_+}} r^{-3}|u|^2\upd r^* \right) \\
&+ \int_0^{R_+} 2g\mathop{\mbox{Re}}(\bar{u}'H)\upd r^* + Q_2^g({R_+})
\end{split}
\label{bounded frequency estimate}
\end{equation}
where the constant $C(\omega_0, \lambda) \sim e^{c \sqrt{\lambda}}$ for some $c$.

\subsubsection{Summing}
We now define the projection of $\phi$ to the angular frequency range $\ell(\ell + 1) \leq \lambda$, which we call $\phi^\flat$, and we similarly define $\phi^\flat_{\hbox{\Rightscissors}}$ etc. Combining the estimates (\ref{almost stationary estimate}) and (\ref{bounded frequency estimate}), taking $R_2 = R_1$, summing over angular frequencies and integrating over time frequencies we obtain
\begin{equation}
\begin{split}
&\int_0^{\tau}\upd t \int\limits_{\Sigma_t\cap \{ r \leq R_1 \} }\left( J^T_\mu[\phi^\flat_{\hbox{\Rightscissors}}]n^\mu_{\Sigma_t} + r^{-2}|\phi^\flat|^2\right) \\ &\lesssim \int_0^{R_1} \upd r^* \int_{-\infty}^\infty \upd\omega \sum_{\substack{0 \leq \ell(\ell + 1) \leq \lambda \\ |m|\leq \ell}} \left(|u'_{m\ell}|^2 + \left(\ell(\ell + 1) + \omega^2 + 1\right)|u_{m\ell}|^2\right) \\
&\lesssim \ e^{c\lambda} \Bigg(R_1^2 \int\limits_{\{ R_1\leq \epsilon_2^{-1}R_1\}\cap\{0\leq t \leq \tau\}}\left(\epsilon_2 r^{-2} + r^{-4}\right)J^T_\mu[\phi_{\hbox{\Rightscissors}}^\flat]T^\mu + R_1^2\int\limits_{\{\epsilon_2^{-1}R_1 \leq r \leq R_+\}\cap\{ 0\leq t \leq \tau\}} r^{-3}J^T_\mu[\phi_{\hbox{\Rightscissors}}^\flat]T^\mu \\
&\phantom{\lesssim} + \int_{-\infty}^\infty \upd\omega \sum_{\substack{0 \leq \ell(\ell + 1) \leq \lambda \\ |m|\leq \ell}} \left(\int_0^{R_+} \left(g\mathop{\mbox{Re}}(\bar{u}_{\ell m}'H_{\ell m}) + h\mathop{\mbox{Re}}(\bar{u}_{\ell m}H_{\ell m})\right) \upd r^* + Q_2^g({R_+})\right)\Bigg)
\end{split}
\label{summed estimate 1}
\end{equation}

Now, we have
\begin{equation}
\begin{split}
& R_1^2 \int\limits_{\{ R_1\leq \epsilon_2^{-1}R_1\}\cap\{0\leq t \leq \tau\}} \left(\epsilon_2 r^{-2} + r^{-4}\right)J^T_\mu[\phi_{\hbox{\Rightscissors}}^\flat]T^\mu + R_1^2\int\limits_{\{\epsilon_2^{-1}R_1 \leq r \leq R_+\}\cap\{ 0\leq t \leq \tau\}} r^{-3}J^T_\mu[\phi_{\hbox{\Rightscissors}}^\flat]T^\mu \\
\lesssim & (\epsilon_2 + R_1^{2} + \epsilon_2^3R_1^{-1}) \int\limits_{\{r \geq R_1\}\cap\{0\leq t \leq \tau\}} J^T_\mu[\phi_{\hbox{\Rightscissors}}^\flat]n_{\Sigma_t}^\mu \\
\lesssim & (\epsilon_2 + R_1^{-2} + \epsilon_2^3R_1^{-1})\int_{\Sigma_0} J^T_\mu[\phi^\flat]n_{\Sigma_0}^\mu 
 + (\epsilon_2R_1^{-1+\delta} + R_1^{-3 + \delta} + \epsilon_2^3R_1^{-2 + \delta})\int\limits_{\{r \geq R_1\}\cap\{0\leq t \leq \tau\}}r^{-1-\delta} J^T_\mu[\phi^\flat] n_{\Sigma_t}^\mu
\end{split}
\end{equation}
where the spatial term arises from estimating the energy in the strips of length $1$ in which the solution $\phi^\flat$ and the cut-off solution $\phi^\flat_{\hbox{\Rightscissors}}$ do not agree. This can be estimated in terms of the initial energy, using the trivial boundedness result (\ref{boundedness}). Note that we plan to absorb the other terms into the left hand side of (\ref{summed estimate 1}), after having added a small amount of our large $r$ current, and chosen the parameters $\epsilon_1$, $\epsilon_2$ and $R_1$ appropriately.

\subsubsection{Errors from the Cut Off}
First some notation: let $\Phi_{\ell m}(\omega, r)$ be the Fourier component of $\phi$, i.e.\ $ u_{\ell m} = r\Phi_{\ell m}$. Then we have
\begin{equation}
\begin{split}
\mathop{\mbox{Re}}(\bar{u}_{\ell m}' H_{\ell m}) &= \mathop{\mbox{Re}}(r\Phi_{\ell m}'H_{\ell m} + \Phi H_{\ell m}) \\
& \leq \epsilon r^{-1-\delta}\left(r^2 |\Phi_{\ell m}'|^2 + |\Phi_{\ell m}|^2 \right) + \frac{1}{\epsilon} r^{1+\delta} |H_{\ell m}|^2
\end{split}
\end{equation}
Similarly, we have
\begin{equation}
\begin{split}
\mathop{\mbox{Re}}(\bar{u}_{\ell m} H_{\ell m}) &= \mathop{\mbox{Re}}(r\Phi_{\ell m}H_{\ell m}) \\
& \leq \epsilon r^{-1-\delta}|\Phi_{\ell m}|^2 + \frac{1}{\epsilon} r^{3+\delta} |H_{\ell m}|^2
\end{split}
\end{equation}
and thus, if we pick any large $R_3$, then
\begin{equation}
\begin{split}
&\int_{-\infty}^\infty \upd\omega \sum_{\substack{0 \leq \ell(\ell + 1) \leq \lambda \\ |m|\leq \ell}} \left(\int_0^{R_3} \left(g\mathop{\mbox{Re}}(\bar{u}_{\ell m}'H_{\ell m}) + h\mathop{\mbox{Re}}(\bar{u}_{\ell m}H_{\ell m})\right)\right) \\
&\leq \int\limits_{\{ r \leq R_3 \}\cap \{0\leq t \leq \tau\}} \epsilon r^{-1-\delta} J^T_{\mu}[\phi^\flat_{\hbox{\Rightscissors}}]n^\mu_{\Sigma_t} + \int\limits_{\{ r\leq R_3\}\cap\{ \{0\leq t\leq 1\}\cup\{\tau -1 \leq t \leq \tau\}\}} \frac{1}{\epsilon} C(R_3) |H|^2 \\
&\leq \epsilon \int\limits_{\{ r \leq R_3 \}\cap \{0\leq t \leq \tau\}} r^{-1-\delta} J^T_{\mu}[\phi^\flat]n^\mu_{\Sigma_t} + \frac{1}{\epsilon} C(R_3)\int_{\Sigma_0}J^T_\mu[\phi^\flat] n^\mu_{\Sigma_0}
\end{split}
\end{equation}
where in the last line we have used our bounds on $H$, (\ref{H bounds}), as well as the compactness of the support of $H$ in time, and also estimated the difference $|J^T_\mu[\phi^\flat_{\hbox{\Rightscissors}}]n^\mu - J^T_\mu[\phi^\flat]n^\mu|$ in terms of the initial data. Note that the first term is of precisely the correct form to be absorbed by our large $r$ current, once $\epsilon$ is chosen small enough.

Now, we could simply take $R_3 = R_+$ in order to bound these terms. However, we wish none of our constants to depend upon $R_+$, since, in order to deal with the boundary term, we will consider compactly supported initial data, and take $R_+$ sufficiently large so that the boundary term vanishes. We then wish to make a density argument to pass back to sufficiently fast decaying initial data, but in order for this argument to work, none of our constants may depend on $R_+$, since we will be taking the limit $R_+ \rightarrow \infty$.

Instead, we choose some $R_3$ such that the functions $g$, $h$ appearing in our currents are identically $1$ and $0$ respectively, for all frequency ranges, in $r \geq R_3$. Then, in $r \geq R_3$, the error term from the cut off is precisely
\begin{equation}
\int_{R_3}^{R_+} \upd r^* \int_{-\infty}^\infty \upd\omega \sum_{\substack{0 \leq \ell(\ell + 1) \leq \lambda \\ |m|\leq \ell}} \mathop{\mbox{Re}}(\bar{u}_{\ell m}' H_{\ell m}) = \int_{\mathbb{S}^2} \upd \Omega \int_{R_3}^{R_+} \upd r^* \int_{-\infty}^{\infty}\upd t \ \partial_{r^*}\left( r\phi^\flat_{\hbox{\Rightscissors}} \right) arH^\flat
\end{equation}
Note that the integrand is actually only supported in the strips $\{0\leq t \leq 1\}\cap\{\tau-1\leq t \leq \tau\}$. Recall that we have the explicit expression for $H^{\flat}$:
\begin{equation}
H^\flat = 2\nabla^\mu \xi_{\tau} \nabla_\mu \phi^{\flat} + (\Box \xi_{\tau})\phi^\flat
\label{explicit cut off}
\end{equation}
From the first term in (\ref{explicit cut off}), we get a term in the integrand of the form
\begin{equation}
\partial_t\xi_{\tau}\left(r^2 \partial_r \phi^\flat_{\hbox{\Rightscissors}}\partial_t\phi^\flat + r \phi^\flat_{\hbox{\Rightscissors}}\partial_t\phi^\flat \right)
\label{first cut off error}
\end{equation}
Once again, we bound the difference in the energy of the cut off solution $\psi^\flat_{\hbox{\Rightscissors}}$ and the solution $\psi^\flat$ using the compactness in time of $H$. The first term in (\ref{first cut off error}) can then be bounded by the initial energy in view of the compact support in time of $\partial_t\xi_{\tau}$. The second term may also be bounded by the same quantity, after employing a Hardy inequality.

We are now faced with bounding a term of the form
\begin{equation}
\begin{split}
&\int_{\mathbb{S}^2} \upd \Omega \int_{R_3}^{R_+} \upd r^* \int_{-\infty}^{\infty}\upd t \ \partial_{r^*}\left( r\phi^\flat_{\hbox{\Rightscissors}} \right) ar (\Box \xi_{\tau})\phi^\flat \\
\end{split}
\end{equation}
Noting that $\Box \xi_{\tau} = -a^{-1}\partial_t^2\xi_{\tau}$, this term is in fact
\begin{equation}
\int_{\mathbb{S}^2} \upd \Omega \int_{R_3}^{R_+} \upd r^* \int_{-\infty}^{\infty}\upd t \ \frac{1}{2}\partial_{r^*}\left(\left(r\phi^\flat \xi_{\tau}\right)^2\right) \xi_{\tau}\partial^2_t \xi_{\tau}
\end{equation}
so we only obtain boundary terms from this error. Using the fact that $\xi_{\tau}\partial^2_t \xi_{\tau}$ is compactly supported in time, the boundary term at $r = R_3$ can easily be bounded by the initial data. On the other hand, the boundary term at $R_+$ will be controlled in the next section.

We have thus obtained
\begin{equation}
\begin{split}
& \int_{-\infty}^\infty \upd\omega \sum_{\substack{0 \leq \ell(\ell + 1) \leq \lambda \\ |m|\leq \ell}} \left(\int_0^{R_+} \left(g\mathop{\mbox{Re}}(\bar{u}_{\ell m}'H_{\ell m}) + h\mathop{\mbox{Re}}(\bar{u}_{\ell m}H_{\ell m})\right)\right) \\
\leq& \epsilon \int_{\{ r \leq R_3 \}\cap \{0\leq t \leq \tau\}} r^{-1-\delta} J^T_{\mu}[\phi^\flat]n^\mu_{\Sigma_t} + \frac{1}{\epsilon} C(R_3)\int_{\Sigma_0}J^T_\mu[\phi^\flat]n^\mu_{\Sigma_0} \\
& + \left(\int_{\mathbb{S}^2} \upd \Omega \int_{-\infty}^{\infty}\upd t \ \frac{1}{2}\partial_{r^*}\left(\left(r\phi^\flat \xi_{\tau}\right)^2\right) \xi_{\tau}\partial^2_t \xi_{\tau}\right)\bigg|_{r = R_+}
\end{split}
\label{boundary cut off error}
\end{equation}

\subsubsection{The Boundary Terms}

The only terms in (\ref{preliminary bounded current}) which we have yet to control are the boundary terms, $Q_2^g({R_+})$ in (\ref{preliminary bounded current}) and the final term in (\ref{boundary cut off error}). To deal with these terms, we first note that, once we have replaced the terms arising from the cut off in time in accordance with the previous subsection, none of the constants depend on the quantity $R_+$. Therefore, if we were to take initial data which is compactly supported, then we could always pick $R_+$ large enough that these boundary terms vanish. We can then use a density argument to remove the dependence on compact support, and instead consider data which is in the closure of the compactly supported initial data with respect to our energy norm, $\int_{\Sigma_0}J^T_\mu[\phi] n^\mu_{\Sigma_0}$. Later, if we wish to obtain pointwise bounds, this norm must be replaced with an appropriate ``higher order energy'' norm.

\subsubsection{Putting it Together}
We have obtained the following estimate:
\begin{equation}
\int_0^{\tau}\upd t \int\limits_{\Sigma_t\cap \{ r \leq R_1 \} }\left( J^T_\mu[\phi^\flat_{\hbox{\Rightscissors}}]n^\mu_{\Sigma_t} + r^{-2}|\phi^\flat|^2\right) \lesssim \epsilon \int\limits_{\{r \geq R_1\}\cap\{0\leq t \leq \tau\}} r^{-1-\delta} J^T_\mu[\phi^\flat]n^\mu_{\Sigma_t} + \int_{\Sigma_0} J^T_\mu[\phi^\flat] n^\mu_{\Sigma_0}
\label{preliminary bounded current 2}
\end{equation}
where we can pick $\epsilon$ as small as we like, and $R_1$ as large as we like. We pick $R_3$ sufficiently large for our previous estimates to hold, and in addition we ensure that $R_3 > R_0$. We now add to this quantity $\epsilon_3$ times the large $r$ current, where $\epsilon_3$ is chosen sufficiently small so that the error term on the right hand side of (\ref{large r equation}) can be absorbed by the left hand side of (\ref{preliminary bounded current 2}). Finally, we choose $\epsilon$ sufficiently small (depending on $\epsilon_3$) in (\ref{preliminary bounded current 2}) so that the error term on the right hand side (the integral in the region $r \geq R_1$) can be absorbed by the contribution from the large $r$ current to the left hand side.

This completes the proof of theorem \ref{ILED theorem}.

\subsection{Weighted Energy}
In this section we will present the ``weighted energy'' estimate which, together with our ILED result \ref{ILED theorem}, will allow us to use the ``black box'' methods of \cite{Dafermos2010b} in order to conclude decay. The weighted energy allows us to gain control in the asymptotic region, whereas our ILED result allows us to control the region near the origin.

Define the null co-ordinates
\begin{equation}
\begin{split}
u &= \frac{1}{2}(t-r^*) \\
v &= \frac{1}{2}(t+r^*)
\end{split}
\end{equation}
in these co-ordinates, the metric is
\begin{equation}
\upd s^2 = -4a \ \upd u \upd v + r^2 \upd \Omega^2
\end{equation}
and the wave equation, $\Box\phi + F\phi = 0$ becomes
\begin{equation}
- \partial_u \partial_v \psi + a \slashed{\triangle}\psi + a\hat{F}\psi = 0
\end{equation}
where, as before, $\psi = r\phi$, and we have defined
\begin{equation}
\hat{F} = F - a^{\frac{1}{2}}b^{-\frac{1}{2}}\partial_r(a^{\frac{1}{2}}b^{-\frac{1}{2}})
\end{equation}

Multiplying by $r^p \partial_v\psi$ and integrating by parts over the spheres, we find
\begin{equation}
\int_{\mathbb{S}^2}\left( r^p \partial_u(\partial_v \psi)^2 + 2a r^p (\slashed{\nabla}\partial_v\psi)\cdot (\slashed{\nabla}\psi) - ar^p \hat{F} \partial_v(\psi^2)\right)\upd\Omega = 0
\end{equation}
Now, we commute $\partial_v$ with $\slashed{\nabla}$, recalling that there is a factor of $1/r$ in $\slashed{\nabla}$:
\begin{equation}
\begin{split}
0 = \int_{\mathbb{S}^2}&\left( r^p \partial_u(\partial_v \psi)^2 + a \partial_v(r^p|\slashed{\nabla}\psi|^2) + (2-p)a^{\frac{1}{2}}b^{-\frac{1}{2}} r^{p-1}|\slashed{\nabla}\psi|^2 - ar^p\hat{F} \partial_v(\psi^2)\right)\upd\Omega \\
= \int_{\mathbb{S}^2}&\left( \partial_u\left(r^p (\partial_v \psi)^2\right) + pr^{p-1} a^{\frac{1}{2}} b^{-\frac{1}{2}} (\partial_v \psi)^2 + \partial_v\left(ar^p|\slashed{\nabla}\psi|^2\right) - a' a^{\frac{1}{2}}b^{-\frac{1}{2}}r^p|\slashed{\nabla}\psi|^2 \right. \\
& + (2-p)a^{\frac{1}{2}}b^{-\frac{1}{2}}r^{p-1}|\slashed{\nabla}\psi|^2 - \partial_v\left(a\hat{F}r^p\psi^2\right) + a^{\frac{1}{2}}b^{-\frac{1}{2}}(a\hat{F})' r^p\psi^2 \\
&\left. + (p+1)a^{\frac{3}{2}}b^{-\frac{1}{2}} \hat{F}r^{p-1}\psi^2 \right)\upd\Omega
\end{split}
\end{equation}

\begin{figure}[ht]
\centering
\includegraphics[scale = 0.6]{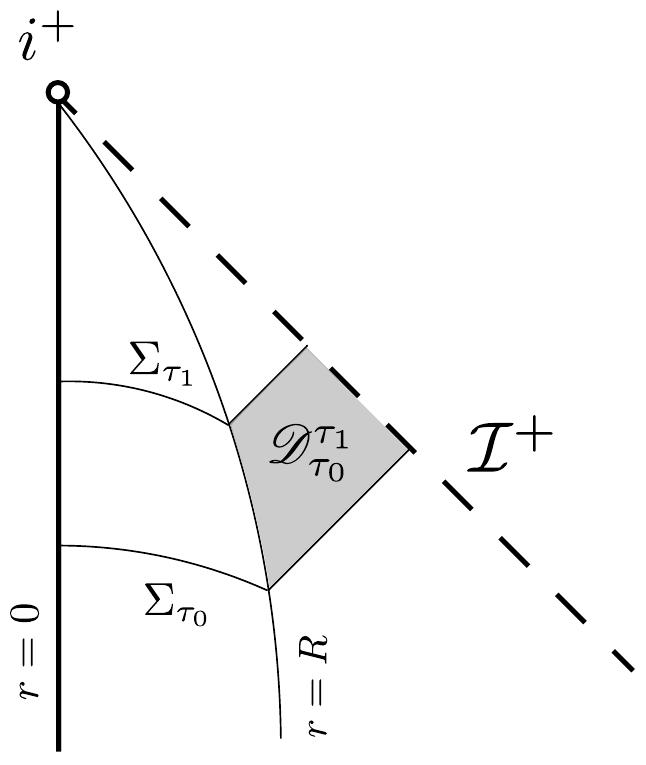}
\caption{The shaded region is where we apply weighted energy estimates. Note that the hypersurfaces $\Sigma_\tau$ contain both a constant $t$ part (in $r \leq R$) and a constant $u$ part (in $r > R$).}
\label{picture1}
\end{figure}

and finally, integrating in $u$ and $v$, in the region $\mathscr{D}_{\tau_0}^{\tau_1}$, which is $r \geq R$, $u_0 \leq u \leq u_1$ (see figure \ref{picture1}) we obtain the following \emph{weighted energy} estimate:
\begin{lemma}
\emph{(Weighted energy estimate)}

Let $\phi$ solve the linear wave equation (\ref{wave equation 1}), and let $\psi = r\phi$. Then:
\begin{equation}
\begin{split}
&\int\limits_{\substack{u=u_1 \\ r\geq R}}\upd v \int_{\mathbb{S}^2} \upd\Omega \ r^p (\partial_v \psi)^2 + \int_{\mathscr{D}_{\tau_0}^{\tau_1}}\bigg( p a^{\frac{1}{2}}b^{-\frac{1}{2}} r^{p-1}(\partial_v \psi)^2 + (2-p)a^{\frac{1}{2}}b^{-\frac{1}{2}}r^{p-1}|\slashed{\nabla}\psi|^2 \\
&\phantom{=} - a^{\frac{1}{2}}b^{-\frac{1}{2}}a'r^p|\slashed{\nabla}\psi|^2 + a^{\frac{1}{2}}b^{-\frac{1}{2}}r^{p-1} \left(r(a\hat{F})' + (p+1)a\hat{F}\right)\psi^2\bigg) \upd u \, \upd v \, \upd \Omega \\
&\phantom{=} + \int\limits_{\substack{\cal{J}^+ \\ u_0 \leq u \leq u_1}}\left( ar^p|\slashed{\nabla}\psi|^2 - a\hat{F}r^p\psi^2\right) \, \upd v \, \upd \Omega \\
& = \int\limits_{\substack{u=u_0 \\ r\geq R}} \upd v \int_{\mathbb{S}^2}\upd\Omega \, r^p(\partial_v\psi)^2 + \int\limits_{\substack{u_0 \leq u \leq u_1 \\ v = R^* - u}}\left( -r^p(\partial_v\psi)^2 + ar^p|\slashed{\nabla}\psi|^2 + a\hat{F}r^p\psi^2 \right)
\end{split}
\label{weighted energy}
\end{equation}
\end{lemma}

\subsection{Energy Decay for Bounded Angular Frequencies}
We now use the method of \cite{Dafermos2010b} to obtain energy decay with respect to the foliation pictured in figure \ref{picture1}. We only outline the argument, as further details can be found in \cite{Dafermos2010b}. First note that, since the wave equation is separable in the angular variables, the projection of the solution onto bounded angular frequencies also satisfies a wave equation:
\begin{equation}
-2\partial_u \partial_v \psi^\flat + a\slashed{\triangle}\psi^\flat + a\hat{F}\psi^\flat = 0
\end{equation}
and so in particular we can appeal to the estimate (\ref{weighted energy}).

We first set $p = 2$ in the weighted energy estimate (\ref{weighted energy}), applied to the solution $\psi^\flat$ in the region between $\Sigma_0$ and $\Sigma_\tau$. After a small amount of averaging of this estimate in $R$, and after using our ILED result \ref{ILED theorem} to control this spacetime integral in terms of the initial energy, we obtain
\begin{equation}
\begin{split}
&\int\limits_{\substack{u=u_\tau \\ r\geq R}}\upd v \int_{\mathbb{S}^2} \upd\Omega \ r^2 (\partial_v \psi^\flat)^2 \\
& \phantom{=} + \int_{\mathscr{D}_{0}^{\tau}}\left( 2 a^{\frac{1}{2}}b^{-\frac{1}{2}} r(\partial_v \psi^\flat)^2 - a^{\frac{1}{2}}b^{-\frac{1}{2}}a'r^2|\slashed{\nabla}\psi^\flat|^2 + a^{\frac{1}{2}}b^{-\frac{1}{2}}r \left(r(a\hat{F})' + 3a\hat{F}\right)(\psi^\flat)^2\right) \upd u \, \upd v \, \upd \Omega \\
& \phantom{=} + \int\limits_{\substack{\cal{J}^+ \\ u_0 \leq u \leq u_\tau}}\left(ar^2|\slashed{\nabla}\psi^\flat|^2 - a\hat{F}r^2(\psi^\flat)^2\right) \, \upd v \, \upd \Omega \\
& \leq \int\limits_{\substack{u=u_0 \\ r\geq R}} \upd v \int_{\mathbb{S}^2}\upd\Omega \, r^2(\partial_v\psi^\flat)^2 + C \int_{\Sigma_0} J^T_\mu[\phi^\flat] n^\mu_{\Sigma_0}
\end{split}
\end{equation}
We now observe that, if we have taken $R$ sufficiently large, then $a' \leq 0$. Also, by the asymptotics conditions on $\hat{F}$, (\ref{asymptotic freedom}), we can absorb the spacetime integral term involving $\hat{F}$ into the term involving $(\partial_v\psi^\flat)^2$ using a Hardy inequality, and additionally the term involving $\hat{F}$ at null infinity vanishes. Thus we find
\begin{equation}
\begin{split}
\int\limits_{\substack{u=u_\tau \\ r\geq R}}\upd v \int_{\mathbb{S}^2} \upd\Omega \ r^2 (\partial_v \psi^\flat)^2 + C\int_{\mathscr{D}_{0}^{\tau}} a^{\frac{1}{2}}b^{-\frac{1}{2}} r(\partial_v \psi^\flat)^2 \upd u \, \upd v \, \upd\Omega &\leq \int\limits_{\substack{u=u_0 \\ r\geq R}} \upd v \int_{\mathbb{S}^2}\upd\Omega \, r^2(\partial_v\psi^\flat)^2 \\
&\phantom{\leq} + C \int_{\Sigma_0} J^T_\mu[\phi^\flat] n^\mu_{\Sigma_0}
\end{split}
\end{equation}
By the pigeonhole principle, we can find a dyadic sequence $\tau_n$ such that
\begin{equation}
\int_{\mathscr{D}_{\tau_n}^{\tau_{n+1}}} a^{\frac{1}{2}}b^{-\frac{1}{2}} r(\partial_v \psi^\flat)^2 \upd u \, \upd v \, \upd\Omega \leq C\tau_n^{-1} \left(\int\limits_{\substack{u=u_n \\ r\geq R}} \upd v \int_{\mathbb{S}^2}\upd\Omega \, r^2(\partial_v\psi^\flat)^2 + C \int_{\Sigma_0} J^T_\mu[\phi^\flat] n^\mu_{\Sigma_0} \right)
\end{equation}

Next, we return to the weighted energy estimate (\ref{weighted energy}), now setting $p = 1$ and applying it in the region between $\Sigma_n$ and $\Sigma_{n+1}$. Once again, we average over $R$ and use our ILED statement, and absorb the terms involving $\hat{F}$ by using a Hardy inequality, to obtain
\begin{equation}
\begin{split}
\int\limits_{\substack{u=u_{n+1} \\ r\geq R}}\upd v \int_{\mathbb{S}^2} & \upd\Omega \ r (\partial_v \psi^\flat)^2 + C_1 \int_{\mathscr{D}_{\tau_{n}}^{\tau_{n+1}}}\left(a^{\frac{1}{2}}b^{-\frac{1}{2}}(\partial_v\psi^\flat)^2 + a^{\frac{1}{2}}b^{-\frac{1}{2}}|\slashed{\nabla}\psi^\flat|^2 \right) \\
&\leq \int\limits_{\substack{u = u_n \\ r\geq R}}\upd v \int_{\mathbb{S}^2}\upd \Omega \, r(\partial_v\psi^\flat)^2 + C_2 \int_{\Sigma_0} J^T_\mu[\phi^\flat]n^\mu_{\Sigma_0} \\
& \leq C\tau_n^{-1} \left(\int\limits_{\substack{u=u_n \\ r\geq R}} \upd v \int_{\mathbb{S}^2}\upd\Omega \, r^2(\partial_v\psi^\flat)^2 + C \int_{\Sigma_0} J^T_\mu[\phi^\flat] n^\mu_{\Sigma_0} \right) + C \int_{\Sigma_0} J^T_\mu[\phi^\flat]n^\mu_{\Sigma_0}
\end{split}
\end{equation}
If we add to this a multiple of our ILED estimate (\ref{ILED equation}), and also observe the identity
\begin{equation}
\int_{v_0}^\infty a^{\frac{1}{2}}b^{-\frac{1}{2}}r^2\left(\partial_v(r\phi)\right)^2 = \int_{v_0}^\infty a^{\frac{1}{2}}b^{-\frac{1}{2}}r^2 (\partial_v\phi)^2 + \int_{v_0}^{\infty}\partial_v\left( a^{\frac{1}{2}} b^{-\frac{1}{2}}r\phi^2\right) - \int_{v_0}^\infty r\phi^2\partial_{r^*}\left(a^{\frac{1}{2}}b^{-\frac{1}{2}}\right)
\end{equation}
the first term of which is comparable to the energy, the second term gives rise to an additional boundary term at $r = R$ which can again be bounded using our ILED statement, and the final term can also be absorbed into the energy after applying a Hardy inequality. Thus, we can obtain
\begin{equation}
\begin{split}
\int_{\tau_n}^{\tau_{n+1}}\upd t \int_{\Sigma_t} J^T_\mu[\phi^\flat] n^\mu_{\Sigma_t} \leq& C\tau_n^{-1} \left(\int\limits_{\substack{u=u_n \\ r\geq R}} \upd v \int_{\mathbb{S}^2}\upd\Omega \, r^2(\partial_v\psi^\flat)^2 + C \int_{\Sigma_0} J^T_\mu[\phi^\flat] n^\mu_{\Sigma_0} \right) \\
& + C \int_{\Sigma_0} J^T_\mu[\phi^\flat]n^\mu_{\Sigma_0}
\end{split}
\end{equation}
and, in view of the trivial energy boundedness statement (\ref{boundedness}), this immediately implies $\tau^{-2}$ decay of the energy along our foliation $\Sigma_t$. This concludes our proof of theorem 
\ref{bounded energy decay theorem}.

\subsection{High Frequencies}

To deal with high frequencies, we establish the following lemma:

\begin{lemma}
The following estimate holds:
\begin{equation}
\int_{\tau_0}^{\tau}\upd t\left( \mathcal{E}^T[\phi^\sharp](t) + \int_{\Sigma_t} \frac{1}{r^2}(\phi^\sharp)^2 \right) \leq \frac{C}{\lambda}\tau\left( \mathcal{E}^T[\phi^\sharp](\tau_0) + \sum_{i = 1}^3 \mathcal{E}^T[\Omega_i \phi^\sharp](\tau_0)\right)
\label{high frequency equation}
\end{equation}
where $\phi^\sharp$ is the projection of $\phi$ onto the angular frequency range $\ell(\ell + 1)\geq \lambda$.
\label{high frequency lemma}
\end{lemma}
\emph{Proof}. By energy boundedness (\ref{boundedness}) we already have that
\begin{equation}
\int_{\tau_0}^{\tau} \mathcal{E}^T[\phi^\sharp](t) \upd t \leq C\tau\left( \mathcal{E}^T[\phi^\sharp](\tau_0) + \sum_{i = 1}^3 \mathcal{E}^T[\Omega_i \phi^\sharp](\tau_0)\right)
\end{equation}
and we now use that the frequencies obey $\lambda \leq \ell(\ell+1)$. So, for example,
\begin{equation}
\frac{\lambda}{r^2}\int_{\mathbb{S}^2} \phi^2 \upd\Omega \leq \int_{\mathbb{S}^2}|\slashed{\nabla}\phi|^2
\end{equation}
so the zeroth order term on the left hand side can be estimated by ${\lambda}(\tau_1 - \tau_0)\mathcal{E}^T[\phi^\sharp](\tau_0)$. Similarly, the first order terms can be estimated by the second term on the right hand side, when we recall the $r$ weights in the angular momentum operators, $\Omega_i \sim r$.

We can now extract a dyadic sequence, $\tau_n$ between $\tau_0$ and $\tau_1$, in which the energy obeys
\begin{equation}
\left( \mathcal{E}^T[\phi^\sharp](\tau_n) + \int_{\Sigma_{\tau_n}} \frac{1}{r^2}(\phi^\sharp)^2 \right) \leq \frac{C}{\lambda}\left( \mathcal{E}^T[\phi^\sharp](\tau_0) + \sum_{i = 1}^3 \mathcal{E}^T[\Omega_i \phi^\sharp](\tau_0)\right)
\end{equation}
and energy boundedness then means that we can remove the dependence on the dyadic sequence.

Note that, in contrast to our decay statement for low angular frequencies (\ref{bounded energy decay equation}), this statement - which is the only tool we will need to control the high angular frequencies - \emph{loses a derivative}. That is to say that the energy at some later time is estimated in terms of a higher order energy (i.e.\ the energy after commuting with angular momentum operators) initially. So, although in our ILED statement we have avoided losing a derivative by restricting to low angular frequencies, we nevertheless are forced to lose a derivative when dealing with high angular frequencies. Indeed, due to the presence of trapping, this is all that can possibly be true, as shown in \cite{Sbierski2013a}.

\subsection{Proof of Logarithmic Decay}
We now have all the ingredients necessary to prove theorem \ref{log decay theorem}. In particular, we have the estimates
\begin{equation}
\begin{split}
\mathcal{E}^T[\phi^\flat](\tau) &\leq C_1 \tau^{-2}e^{C_2 \sqrt{\lambda}}\left(\mathcal{E}^T[\phi^\flat](0) + \int\limits_{\substack{u = u_0 \\ r\geq R}}\upd v \int_{\mathbb{S}^2}\upd\Omega \, r^2 (\partial_v\psi^\flat)^2 \right) \\
\left( \mathcal{E}^T[\phi^\sharp](\tau) + \int_{\Sigma_{\tau}} \frac{1}{r^2}(\phi^\sharp)^2 \right) &\leq \frac{C}{\lambda}\left( \mathcal{E}^T[\phi^\sharp](0) + \sum_{i = 1}^3 \mathcal{E}^T[\Omega_i \phi^\sharp](0)\right)
\end{split}
\label{both estimates}
\end{equation}
adding these two estimates, using a Hardy inequality and estimating the angular frequency localised estimates, which appear on the right hand side of (\ref{both estimates}), by the corresponding quantities without angular frequency localisation, we obtain
\begin{equation}
\begin{split}
\mathcal{E}^T[\phi](\tau) + \int_{\Sigma_{\tau}}\frac{1}{r^2}\phi^2 \leq C_1&\Bigg(\left( \tau^{-2}e^{C_2\sqrt{\lambda}} + \frac{1}{\lambda}\right)\mathcal{E}^T[\phi](0) \\
& + \tau^{-2}e^{C_2\sqrt{\lambda}}\int\limits_{\substack{u = u_0 \\ r\geq R}}\upd v \int_{\mathbb{S}^2}\upd\Omega \, r^2 (\partial_v\psi^\flat)^2 + \frac{1}{\lambda}\sum_{i = 1}^3 \mathcal{E}^T[\Omega_i \phi](0)\Bigg)
\end{split}
\label{almost finished estimate}
\end{equation}
The key observation now is that $\lambda > 0$ can be chosen arbitrarily, and in particular we can choose 
\begin{equation}
\lambda = \frac{1}{(C_2)^2}(\log\tau)^2
\end{equation}
which implies $(\log\tau)^{-2}$ decay, and proves theorem \ref{log decay theorem}.

Note that, if one defines higher order energies by commuting with $T$ and the $\Omega_i$, then the factors of $\lambda^{-1}$ on the right hand side of (\ref{almost finished estimate}) can be improved to $\lambda^{-n}$, which leads to decay with a rate $(\log \tau)^{-n-1}$. This should be compared with our remarks at the end of section \ref{log decay lower bound section}. Note also that, since the angular momentum operators contain a weight, this increased decay rate comes at the cost of demanding more localisation of the initial data, as well as higher regularity.

\subsection{Pointwise Decay}
We can now commute once more with the angular derivative operators $\Omega_i$ and use the Sobolev inequality on the spheres to obtain pointwise decay, proving corollary \ref{pointwise decay corollary}.

\subsection{Discussion}
The results of this section are of less directly physical consequence - as explained in the introduction, once we have shown logarithmic decay as a \emph{lower bound}, we already expect that the full nonlinear problem of gravitational waves impinging on an ultracompact neutron star exhibits an instability. However, it should be noted that the results of both this section and the preceding one actually apply to a far wider class of linear wave equations on a wider class of backgrounds than just ultracompact stars. Even when we specialise to such neutron stars, the results of section \ref{lower bound section} could be viewed as obsolete, if, for example, there were to exist exponentially growing mode solutions. On the other hand, the results of this section show that, if an instability is present, then it must be of the more subtle character which we have proposed.

An important aspect of the results of this section are their robustness. Since we neither have an explicit form of the metric, nor the function $F$ which appears in the wave equation (\ref{wave equation 1}), we have had to use very robust methods to obtain our results. As such, they apply to a very wide class of spacetimes and linear wave equations, so long as we exclude ``long range'' forces (by ensuring that $F$ falls off sufficiently quickly) and restrict to asymptotically flat spacetimes. Within this class of spacetimes, the major restriction on the results is once again due to spherical symmetry, as well as to manifolds with trivial topology.

We can contrast this situation with the results of section (\ref{lower bound section}). Whilst the results concerning the lower bound are independent of both the topology of the manifold and its asymptotic properties, the results concerning the upper bound seem to rely on both of these. In fact, while the asymptotic properties are important, it is fairly easy to modify our results to apply, for example, to an asymptotically flat black hole.

Both the results of this section and the previous section do, however, rely upon spherical symmetry. Whilst the results of the previous section rely on symmetry assumptions in a seemingly fundamental way - they depend upon the separability of the wave equation - those of this section are actually far more flexible. Indeed, the only times when we make explicit use of spherical symmetry in this section is when we commute with the angular momentum operators $\Omega_i$. An alternative is to commute with the $\Omega_i$ at large $r$, and some combination of $T$, $\Omega_i$ and $Y$ at smaller $r$, and use elliptic estimates (see, for example, \cite{Dafermos2013} or \cite{Dafermos2014}). Thus, the spherical symmetry is not such a fundamental issue when seeking to generalise our proof of the \emph{upper} bound. It turns out that the major obstacle is in fact the nondegenerate energy bound (\ref{boundedness}), which is trivial in the case of spherical symmetry, but is generally much harder for a rotating spacetime due to the phenomenon of \emph{superradiance}. This is made manifest in the failure of the natural, $T$-energy $\mathcal{E}^T$ to be positive definite. In the case of a spacetime without a boundary this is expected to be fatal (see \cite{Friedman1978}). By contrast, on a black hole background one has the additional tool of the \emph{redshift} vector field. For slowly rotating, non-extremal black holes, for example, one could attempt to exploit this effect in order to recover our results.

In this section we have proven logarithmic decay for linear waves on spherically symmetric spacetimes, under very general conditions. The results of the previous section show that, if the spacetime also exhibits stable trapping, then this is the fastest uniform rate one can achieve, which is highly suggestive of a nonlinear instability. Whilst these results are very general, in section \ref{neutron star section} we have also provided a particular example of an interesting spacetime to which our results apply, which suggests a new lower bound ($r > 3M$) on the radius of (stable) fluid stars, independent of their equation of state.

\section*{Acknowledgements}
 The author is very grateful to Harvey Reall for suggesting this project, and the application of the ideas to ultracompact stars, and also to Gustav Holzegel for helpful comments on the manuscript. The author wishes to thank Mihalis Dafermos and Princeton University for their kind hospitality for part of the duration of this project, and Yakov Shlapentokh-Rothman for many helpful discussions. This work was supported by the European Research Council Grant No. ERC-2011-StG 279363-HiDGR.

\sloppy
\printbibliography

\end{document}